%% file: GAMAautomaticclassification.tex
\documentclass[twocolumn,useAMS,usenatbib]{mnras}
\pdfoutput=1
\usepackage{graphicx}
\usepackage{multirow,bigdelim}
\usepackage{lipsum}
\usepackage{lmodern}
\usepackage{hyperref}
\usepackage{titlesec}
\usepackage{color}
\usepackage{tabularx}
\usepackage{chngcntr}
\usepackage{endnotes}
\usepackage{lscape}
\usepackage{afterpage}
\usepackage{adjustbox}
\usepackage{blindtext}
\usepackage{rotating}
\usepackage[singlelinecheck=false]{caption}

\usepackage[T1]{fontenc}
\usepackage[latin9]{inputenc}
\usepackage[authoryear]{natbib}
\makeatletter
\pdfpageheight\paperheight
\pdfpagewidth\paperwidth
\usepackage[final]{pdfpages}
\usepackage{graphicx,amsmath,amssymb} 
\renewcommand\[{\begin{equation}}
\renewcommand\]{\end{equation}}
\usepackage{aas_macros}

\def\lesssim{\mathrel{\hbox{\rlap{\hbox{\lower4pt\hbox{$\sim$}}}\hbox{$<$}}}}
\def\gtrsim{\mathrel{\hbox{\rlap{\hbox{\lower4pt\hbox{$\sim$}}}\hbox{$>$}}}}

\makeatother
\usepackage{color,xcolor,mathrsfs,braket,pdflscape}

\usepackage{epstopdf}
\usepackage{float}
\input{comms.tex}
\title[GAMA: Automatic Morphological Classification of Galaxies]
{Galaxy And Mass Assembly: Automatic Morphological Classification of Galaxies Using Statistical Learning}
\author[S.Sreejith et al.]{Sreevarsha Sreejith,$^{1}$
Sergiy Pereverzyev Jr.,$^{2}$
Lee S. Kelvin,$^{1,3}$
Francine Marleau,$^{1}$
\newauthor
Markus Haltmeier,$^{2}$
Judith Ebner,$^{2}$
Joss Bland-Hawthorn,$^{4}$
Simon P. Driver,$^{5,6}$
\newauthor
Alister W. Graham,$^{7}$
Benne W. Holwerda,$^{8}$
A. M. Hopkins,$^{9}$
J. Liske,$^{10}$
\newauthor
Jon Loveday,$^{11}$
Amanda J. Moffett,$^{12}$
K. A. Pimbblet,$^{13,14}$
Edward N. Taylor,$^{7}$
\newauthor
Lingyu Wang,$^{15,16}$
Angus H. Wright$^{17}$
\\
\\
\\
$^{1}$Institute for Astro-and Particle Physics, University of Innsbruck, Austria\\
$^{2}$Department of Mathematics, University of Innsbruck, Austria\\
$^{3}$Astrophysics Research Institute, Liverpool John Moores University, IC2, Liverpool Science Park, 146 Brownlow Hill, Liverpool, L3 5RF, UK\\
$^{4}$Sydney Institute for Astronomy, School of Physics A28, University of Sydney, NSW 2006, Australia\\
$^{5}$International Centre for Radio Astronomy Research (ICRAR), University of Western Australia, Crawley, WA 6009, Australia\\
$^{6}$Scottish Universities' Physics Alliance (SUPA), School of Physics and Astronomy, University of St Andrews,\\ North Haugh, St Andrews, KY16 9SS, UK\\
$^{7}$Centre for Astrophysics and Supercomputing, Swinburne University of Technology, Victoria 3122, Australia.\\
$^{8}$Department of Physics and Astronomy, University of Louisville, Louisville KY 40292, USA\\
$^{9}$Australian Astronomical Observatory, PO Box 915, North Ryde, NSW 1670, Australia\\
$^{10}$Hamburger Sternwarte, Universit{\"a}t Hamburg, Gojenbergsweg 112, 21029 Hamburg, Germany\\
$^{11}$Astronomy Centre, University of Sussex, Falmer, Brighton BN1 9QH, UK\\
$^{12}$Department of Physics \& Astronomy, Vanderbilt University, Nashville TN 37240, USA\\
$^{13}$E.A.Milne Centre for Astrophysics, University of Hull, Cottingham Road, Kingston-upon-Hull, HU6 7RX, UK\\
$^{14}$School of Physics and Astronomy, Monash University, Clayton, VIC 3800, Australia\\
$^{15}$SRON Netherlands Institute for Space Research, Landleven 12, 9747 AD, Groningen, The Netherlands\\
$^{16}$Kapteyn Astronomical Institute, University of Groningen, Postbus 800, 9700 AV, Groningen, The Netherlands\\
$^{17}$Argelander Institut f{\"u}r Astronomie, Universit{\"a}t Bonn, Auf dem H{\"u}gel 71, 53121 Bonn, Germany
}
\date{\today}
\begin{document}
\maketitle
\begin{abstract}

We apply four statistical learning methods to a sample of $7941$ galaxies ($z<0.06$) from the Galaxy and Mass Assembly (GAMA) survey to test the feasibility of using automated algorithms to classify galaxies. Using $10$ features measured for each galaxy (sizes, colours, shape parameters \& stellar mass) we apply the techniques of Support Vector Machines (SVM), Classification Trees (CT), Classification Trees with Random Forest (CTRF) and Neural Networks (NN), returning True Prediction Ratios (TPRs) of $75.8\%$, $69.0\%$, $76.2\%$ and $76.0\%$ respectively. Those occasions whereby all four algorithms agree with each other yet disagree with the visual classification (`unanimous disagreement') serves as a potential indicator of human error in classification, occurring in $\sim9\%$ of ellipticals, $\sim9\%$ of Little Blue Spheroids, $\sim14\%$ of early-type spirals, $\sim21\%$ of intermediate-type spirals and $\sim4\%$ of late-type spirals \& irregulars. We observe that the choice of parameters rather than that of algorithms is more crucial in determining classification accuracy. Due to its simplicity in  formulation and implementation, we recommend the CTRF algorithm for classifying future galaxy datasets. Adopting the CTRF algorithm, the TPRs of the 5 galaxy types are : E, $70.1\%$; LBS, $75.6\%$; S0-Sa, $63.6\%$; Sab-Scd, $56.4\%$ and Sd-Irr, $88.9\%$. 
 Further, we train a binary classifier using this CTRF algorithm that divides galaxies into spheroid-dominated (E, LBS \& S0-Sa) and disk-dominated (Sab-Scd \& Sd-Irr), achieving an overall accuracy of $89.8\%$. This translates into an accuracy of $84.9\%$ for spheroid-dominated systems and $92.5\%$ for disk-dominated systems.
\end{abstract}

{\bf Keywords:} galaxies, automated morphological classification, statistical learning, machine learning algorithms.

\section{Introduction}

\label{sec:intro}Galaxies are observed to have a wide variety of forms,
from bright massive ellipticals to extended late-type spirals and 
faint compact dwarfs. One of the first attempts in categorising galaxies
by their visual appearance was proposed by \citet{Wolf1908}. These
so-called `galactic nebulae' were arranged according to their shape,
size and distinguishing features. No continuity or transition
between these groupings was suggested. As imaging technology improved
over the course of the next decade and available datasets grew, new
systems for galaxy classification were proposed by many authors (e.g.:
\citealp{Jeans1919,Reynolds1920}). This 
culminated in the development of the \citet{Hubble1936a} sequence or tuning fork. The Hubble tuning fork divides
galaxies into early type\footnote{The naming conventions `early type' and `late type' refer to the complexity of visual appearance, and do not imply (nor was it meant to imply) an evolutionary sequence \citep{Baldry2008b}.}: typically red and smooth ellipticals; late type: typically blue extended disk-like spirals, both barred and unbarred, and; a bridging
population of lenticulars: systems with both a smooth bulge component
and an extended yet smooth disk component. Subsequent extensions to
the Hubble tuning fork have addressed a number of shortcomings in
the initial classification methodology. These include the inclusion
of bulge-less spirals \citep{Shapley1940}, transition lenticulars
\citep{Holmberg1958}, rings \citep{deVaucouleurs1959}, barred lenticulars
\citep{Sandage1961,Sandage1975} and dwarfs/irregulars \citep{Sandage1984}.
The success of this relatively simple and extensible schema for morphological
classification of galaxies has ensured that the Hubble tuning fork
remains relevant almost a century later.

Hubble type classifications have been used to explore a number of
astrophysical phenomena. It was initially noted by \citet{Hubble1931}
that elliptical and lenticular galaxies preferentially favour galaxy
cluster environments, indicating a potential environmental dependence
on galaxy morphology. \citet{Oemler1974} built upon this work some
decades later, showing that the early type galaxy fraction increases
in dense regions. \citet{Dressler1980} conclusively showed how the
fractions of elliptical, lenticular and spiral+irregular galaxies
varied as a function of projected galaxy density: the morphology-density
relation. He found that dense regions such as galaxy groups and clusters
preferentially harbour elliptical galaxies, whilst less dense `field'
regions host lenticular, spiral and irregular galaxies (See also \citealt{Smith2005}). This apparent
relation between morphology and environment has been further explored
in recent years to encompass, amongst others, galaxy mass \citep{VanderWel2008},
star formation \citep{Welikala2008,Welikala2009}, colour \citep{Bamford2009},
the galaxy luminosity function (\citealp{Kelvin2014a}, see also \citealp{Baldry2006}),
the galaxy stellar mass function \citep{Kelvin2014b} and galaxy structure
\citep{Hiemer2014}. 

Precisely how galaxies form and evolve into their various morphological
configurations, and the dependence of this on environment, has been
the subject of much investigation. \citet{Spitzer1951} first suggested
that merging events between galaxies, more common in dense cluster
environments, may be responsible for their transition from a spiral
to a lenticular morphology. \citet{Toomre1977} went further, suggesting
that elliptical galaxies may also be formed via this merging mechanism
(see also \citealp{White1978}). In addition to merging, a number
of supplementary processes which act to modify the morphology of a
galaxy have been proposed, including ram pressure stripping of spiral
gas as a galaxy travels through a hot dense intracluster medium \citep{Gunn1972},
the rapid decline of star-formation due to a loss of its hot gas resevoir
\citep[strangulation:][]{Larson1980,Kauffmann1993,Balogh2000,Diaferio2001},
heating of the galaxy caused by rapid encounters with other nearby
systems \citep[harassment:][]{Moore1996} and tidal interations \citep{Moss2000,Gnedin2003a,Gnedin2003b,Park2008}.
Obtaining an accurate estimate of galaxy morphology is therefore essential
in order to facilitate exploration of the formation and evolution
of galaxies.

Contemporary catalogues of galaxy morphology vary in size and classification
methodology. \citet{Kelvin2014a} (also \citealp{Moffett2016}) morphologically
classify a local volume-limited sample of galaxies taken from
the Galaxy And Mass Assembly (GAMA\footnote{http://www.gama-survey.org},
\citealp{Driver2009}) survey. Classification is performed via majority
observer consensus based on visual inspection of a composite three-colour
optical-NIR image. Three independent expert classifiers are asked
a series of questions for each galaxy: is the galaxy spheroid or disk
dominated, is the galaxy a single or multi-component system, and is
the galaxy barred or unbarred. This allows for the galaxy sample to
be principally divided into elliptical (E), early-type spiral (S0-Sa),
intermediate-type spiral (Sab-Scd) and late-type spiral/irregular
(Sd-Irr). Additional barred classes for early-type and intermediate-type
spirals (SB0-SBa and SBab-SBcd, respectively) are also present. A
small subset of `little blue spheroid' (LBS) galaxies, blue compact systems
($\sim7.4$\%), did not fit into this classification hierarchy and
were excluded at the top level. This methodology produces accurate
classifications yet remains a time consuming exercise, a problem which
will only become more acute as future datasets increase in size.

A novel alternative is to enlist the support of the wider astronomy
community. The Galaxy Zoo project \citep{Lintott2008} allows for
volunteer `citizen scientists' to visually classify galaxies via a
web interface. The simple and effective design of the website allows
for a large number of classifiers to visit each galaxy (typically
of the order $\sim60$), enabling rapid classification of large datasets.
However, future facilities such as the Euclid space telescope and Large Synoptic Survey Telescope (LSST) will probe
much larger volumes, providing datasets for several billion galaxies.
For these future facilities, morphological classification via visual
inspection becomes increasingly prohibitive.

The concept of using automated techniques to quantify galaxy morphologies stem from this `big data overload' scenario. \citet{Moore2006} demonstrated the use of an automated Mathematical Morphology algorithm to achieve classification into ellipticals and late-type spirals using the images from \citet{Smail1997}. Their approach was unique in that it had fewer free parameters and that it didn't require a classifier to be trained with a machine learning algorithm. Another widely used approach to classify galaxies is by the application of statistical machine learning algorithms. Those that have been used previously used include artificial neural networks, Support Vector Machines (SVM), decision trees and random forests. They are applied to either galaxy images or to parameters extracted from imaging and spectroscopic data. As part of the Kaggle challenge conducted by the Galaxy Zoo team, \citet{Dieleman2015} presented a convolutional neural network approach (ConvNets) to classify galaxy images. Their algorithm was designed to operate with a training set of $55,420$ galaxy images,  real time evaluation set of $6158$ images and a test set of $79,975$ images. \citet{Huertas-Company2015} applied this algorithm to $58,000$ (47,700 training, 5300 validation and 5000 testing) high redshift galaxy images\footnote{The training set actually consists of 8000 galaxies from the GOODS-S field which are rotated randomly three times and over three filters to obtain 58000 galaxy images \citep{Huertas-Company2015}.}(median redshift $z\sim1.25$) from 5 Cosmic Assembly Near-infrared Deep Extragalactic Legacy Survey (CANDELS) fields with a result of $<1\%$ misclassifications.

\citet{abraham1996}\footnote{The use of concentration index parameter for galaxy classification can be traced as far back to \citet{Shapley1927} and \citet{Morgan1958}.} introduced a new method of discerning between early, late and irregular type galaxies, the C-A plane, where C stands for the central concentration and A for the rotational asymmetry of the galaxy. This was based on \citet{Okamura1984} and \citet{Doi1993}, both of whom proposed a strong correlation between the mean concentration index and galaxy morphology. The logged values of these two parameters are plotted in a 2-D plane and the separation between the different galaxy populations are obtained by applying linear boundaries. \citet{conselice2003} expanded upon this method by adding a third dimension, smoothness or clumpiness of the galaxy (represented by S). He was also among the first groups to consider additional morphological types such as dwarf ellipticals, dwarf irregulars and mergers. For more than three dimensions \footnote{Please note that dimensions refer to the number of parameters used for the classification process. This terminology is used increasingly when referring to SVM methods where a kernel function (Gaussian in most cases) is applied to non linearly separable data to project the parameter space into a higher dimension where the data are linearly separable.}, this method becomes difficult. Also, it presents some problems when it comes to ground based, high redshift data. \citet{Graham2001b} revealed that the concentration parameter, C was unstable in nature due to its high sensitivity to the image exposure depth. \citet{conselice2003} explains that while it is possible to obtain average values for CAS parameters for data from space-based telescopes (deep Hubble Space Telescope data being the example in the paper) up to a redshift $z\sim3$, the same values for single galaxies will have such high uncertainties that their usage will be quite limited until such a time when deeper and high resolution imaging can be taken. 

 \citet{Huertas-Company2007} offered a generalisation of the CAS method using SVM. Other examples from literature where a statistical learning technique was used to classify galaxies include \citet{banerji2010} (artificial neural networks), \citet{owens1996} (oblique decision trees) and \citet{gauci2010} (three decision tree algorithms including a random forest approach). All these methods use measured parameters as inputs to the classifying algorithms.

The goal of this paper is to explore the viability in using statistical
learning methods to produce robust automated Hubble-type morphology
catalogues for datasets with a greater variety in galaxy types. We have attempted to formulate a general method that will be applicable to small data sets and surveys that do not have access to such a wide variety of parameters as we do. Section 2 details the GAMA \citep{Driver2009} dataset used in this study.
Section 3 describes the various statistical learning algorithms under
consideration and the application of these algorithms
to the dataset. Results are shown in Section 4 and the conclusions
and future prospects are presented in Section 5. Unless otherwise
stated, a standard cosmology of $\left(H_{0},\Omega_{m},\Omega_{\Lambda}\right)=\left(70\mathrm{\,km\,s^{-1}\,Mpc^{-1}},0.3,0.7\right)$
is assumed throughout this paper.

\section{Data}
In this section we briefly describe the GAMA survey from which our
data sample is taken, the parameters that we have chosen and the
justifications for choosing these specific parameters.

\subsection{Galaxy And Mass Assembly (GAMA) }

GAMA is a project designed to study the low redshift galaxy population, combining data from eight ground-based and four space-based facilities. It involves both spectroscopic and multi-wavelength imaging programmes which are designed to study
structures along the scales from 1 kiloparsec (kpc) to 1 megaparsec (Mpc) in the nearby
 Universe ($z\lesssim0.25$). The main goal of the GAMA survey is to test and verify the
hierarchical structure formation scenario that emerges from the
$\Lambda$CDM cosmological model by measuring the structure growth
rate, halo mass function and star forming efficiency of galaxies in
groups.

The GAMA spectroscopic survey was carried out on the AAOmega
multi-object spectrograph on the Anglo-Australian Telesecope (AAT). It includes $\sim300,000$ galaxies with magnitudes down to r $\sim19.8$ mag
(r being the Galactic extinction corrected Petrosian magnitude in the
r-band from SDSS DR6; \citealt{AdelmanMcCarthy2008}) spanning an area of $\sim286$ deg$^2$. The GAMA imaging programme compiles and reprocesses data from a number of other contemporary imaging surveys (see \citealt{Driver2009} for details). The reprocessed optical and near-infrared imaging has a pixel-scale resolution of $0.339$ arcseconds/pixel. The master GAMA input catalogue, InputCatAv07, is primarily based on SDSS DR7 \citep{Abazajian2009} photometry. The majority of the redshifts have been attained as part of the GAMA spectroscopic campaign on the AAT \citep{Hopkins2013}. Additional redshifts are obtained from a number of surveys including the SDSS \citep{smee2013}, 2dFGRS \citep{colless2001}, MGC \citep{Driver2005} and others. Full details may be found in \citet{Driver2009} and \citet{Baldry2014}. 

\begin{table*} 
\caption{Hubble type classifications in the GAMA Catalogue and their distribution in our data set. The complete data set consists of 7941 objects from which we remove 374 objects that are visually classified as a `star' or `artefact' (GAMA Hubble types 50 and 60) and 39 objects that do not have valid values for the parameters we have chosen. Of the remaining 7528 objects, we combine the unbarred (11) and barred (12) early type spirals as well as the unbarred (13) and barred (14) intermediate-type spirals to form two new composite data types 1112 and 1314 (henceforth combinedly referred to as S0-Sa and Sab-Scd respectively).}
\label{gamatypes}
\resizebox{\textwidth}{!}
{\begin{tabular}{llllll}

\hline
\ct{3cm}{GAMA \\Hubble\\type code}  & Galaxy type  & Abbreviation & \ct{3cm}{Number of objects \\(\% in final \\7528 sample)} & \ct{3cm} {Of which\\in training set} & \ct{3cm} {Of which \\ in test set} \\
 \hline
\vspace{0.1cm}
1 & Elliptical & E & 856 ($11.4\%\pm3.3$) & 682 (11.3\%) & 174 (11.6\%)\\
\vspace{0.5cm}
2 & Little Blue Spheroid &LBS& 869 ($11.5\%\pm2.0$) & 689 (11.4\%) & 180 (12.0\%) \\
\vspace{0.1cm}

11 & Early-type spirals & S0-Sa \hspace{0.3cm}\rdelim\}{3}{12pt} & \multirow {2} {*} {{833 ($11.1\%\pm0.7$)}} & \multirow {2} {*} {{657 (10.9\%)}} & \multirow {2} {*} {{ 176 (11.7\%)}}\\

\vspace{0.5cm}
12 & \ct{3cm}{Early-type spirals\\(barred)} & SB0-SBa \\
\vspace{0.1cm}
13 & Intermediate-type spirals & Sab-Scd \hspace{0.3cm}\rdelim\}{3}{12pt}& \multirow {2} {*} {{1432 ($19.0\%\pm6.2$)}} & \multirow {2} {*} {{1152 (19.1\%)}} & \multirow {2} {*} {{ 280 (18.6\%)}}\\
\vspace{0.5cm}
14 & \ct{3cm}{Intermediate-type \\spirals (barred)} & SBab-SBcd \\
\vspace{0.5cm}
15 & Late-type spirals \& Irregulars & Sd-Irr &3538 ($47.0\%\pm5.9$)&2842 (47.2\%)& 696 (46.2\%)\\
\vspace{0.1cm}

50 & Artefact & Artefact \hspace{0.3cm}\rdelim\}{3}{12pt} & \multirow {2} {*} {{374 }} & \multirow {2} {*} {{-}} &\multirow {2} {*} {{-}} \\

\vspace{0.5cm}
60 & Star  & Star \\
\vspace{0.5cm}
-& Incomplete features &- & 39& -&- \\
\hline

\end{tabular}}
\raggedright Note: Additional Hubble types of Not Elliptical ($10$) and Uncertain ($70$) Morphologies are available in the GAMA VisualMorphology DMU, though these were derived for a different sample via a different method and as such are not used in this study (see \citealt{driver2012} for further details).
\end{table*}

\begin{table*}
\caption{Parameters chosen from the GAMA catalogues and the derived parameters used for training and testing our algorithms. The parameters in the top panel are those given to the machine learning algorithms as input. Those in the bottom panel are used to derive those in the top panel (with the exception of visual Hubble type), but were not used directly.}
\label{features}

\resizebox{\textwidth}{!}
{\begin{tabular}{llllll}
\hline
\ct{3cm}{Parameter\\ Name}  & \ct{3cm}{Catalogue\\ column name} & Notes & Units & Table & Reference \\

\hline

Stellar mass &logmstar &\ct{3cm}{ logged in \\catalogue }&  $\log_{10}(M_\odot)$&StellarMassesv18 & \citet{taylor2011}\\

 Mass-to-light ratio &      logmoverl\_i &\ct{3cm}{logged in \\catalogue} &$\log_{10}(M_\odot/L_{\odot,i}) $& StellarMassesv18 & \citet{taylor2011}\\
 
 \\$g-i$ colour &      gminusi &  not logged &  mag &   StellarMassesv18 & \citet{taylor2011}\\

 \\$u-r$ colour &      uminusr & not logged & mag &StellarMassesv18 & \citet{taylor2011}\\

 \\Absolute magnitude &     absmag\_r & not logged  & mag &StellarMassesv18 & \citet{taylor2011} \\

\\Ellipticity &      GALELLIP\_r &  not logged  &no unit &SersicCatSDSSv09 & \citet{kelvin2012}\\

 \\S\'ersic index & GALINDEX\_r  & logged & no unit & SersicCatSDSSv09 & \citet{kelvin2012}\\

 \\ \ct{3cm}{Half-light radius\\ in kpc} & - & logged & $\log_{10}(kpc)$& - & - \\
 \\ \ct{3cm}{Kron radius \\ in kpc\\(semi-major axis)}& - & logged & $\log_{10}(kpc)$& - & - \\
\\ \ct{3cm}{Kron radius \\in kpc\\(semi-minor axis)}& - & logged & $\log_{10}(kpc)$& - & - \\

\hline

 \\Half-light radius & GALRE\_r & -  &arcsec &SersicCatSDSSv09 & \citet{kelvin2012} \\

 \\ Kron radius &  KRON\_RADIUS & - & \ct{3cm}{units of A\_IMAGE \\or B\_IMAGE} & ApMatchedCatv06    & \citet{hill2011}\\

\\ \ct{3cm}{Angular size  \\ (semi-major axis) } & A\_IMAGE & \ct{3cm}{used to calculate\\ Kron radius in kpc} & pixels & ApMatchedCatv06    & \citet{hill2011}\\

 \\ \ct{3cm}{Angular size\\ (semi-minor axis)} & B\_IMAGE & \ct{3cm}{used to calculate \\Kron radius in kpc} & pixels & ApMatchedCatv06    & \citet{hill2011}\\

\\Redshift & Z\_TONRY & \ct{3cm}{used to calculate \\Kron and half-light radii in kpc} & no unit & DistancesFramesv14 & \citet{Baldry2012}\\

\\Hubble type \ &  HUBBLE\_TYPE\_CODE & \ct {3cm}{barred and unbarred counterparts merged for training the algorithms} & no unit & VisualMorphologyv02 &\ct{3cm}{ \citet{Kelvin2014a} \\ \citet{Moffett2016}}\\
\\
\hline
\end{tabular}}

\end{table*}

\subsection{Galaxy Sample}\label{s_galsam}

The galaxy sample used in this paper is from Data Release 2 of the
GAMA survey \citep{liske2015} which gives spectra,
redshifts and supplementary information regarding $72,225$ objects from
GAMA Data Release 1 \citep{driver2011}. Our primary sample
consists of $7941$ galaxies which have been visually classified into $11$
Hubble types (\citealt{Kelvin2014a}, \citealt{Moffett2016} ; see Table~\ref{gamatypes}; refer to the VisualMorphologyv02 catalogue in the VisualMorphology DMU for further details), spanning a redshift range of $0.002 \le z \le 0.06$.

From our intial sample of $7941$ galaxies, we have excluded those objects that are classified as a `star' or `artefact' (GAMA Hubble type codes $50$ and $60$; 374 in number) in the VisualMorphology02 catalogue. We have also excluded an additional $39$ objects for which the values were missing for one or more of our chosen parameters. Therefore the final sample that we apply our statistical learning methods to consists of $7528$ objects. Of these, the number of objects of each morphological type are : ellipticals - 856 ($11.4\%\pm3.3$), LBS - 869 ($11.5\%\pm2.0$), early-type spirals - 833 ($11.1\%\pm0.7$), intermediate-type spirals - 1432 ($19.0\%\pm6.0$) and late-type spirals \& irregulars - 3538 ($47.0\%\pm5.9$). We computed uncertainties in the sample based on standard deviations of the classifications by the three human classifiers.

\subsection{Chosen Parameters}\label{s_param}

The choice of input parameters is crucial for the effectiveness of statistical learning algorithms. We want to recreate the classification process that the human eye would perform upon seeing an image, using parameters extracted from such an image. Ideally we would choose parameters that clearly demarcate the different classes of galaxies. Table~\ref{features} lists the parameters that we have chosen from the GAMA database  for each galaxy, the tables they have been taken from and the relevant references. 

It is decidedly non-trivial to differentiate between galaxies using only parameters that give similar information, for example, galaxy colour. In \citet{lange2015}, the separation between early and late type galaxies in the GAMA catalogue are defined as $u-r = 1.5$ mag and $g-i = 0.65$ mag. Values greater than these would represent the redder (early-type) galaxies while values less than these would represent bluer (late-type) galaxies. Using only colour to ascribe morphology of a galaxy gives a good general picture of the apparent bimodality of the local galaxy population, but neglects the fact that colour traces star formation while morphology reflects the dynamic evolution of the galaxy. While they are related, they are not the same. The colour information alone may bias against certain morphological types such as blue ellipticals and red spirals (see Figure 20 of \citealt{kelvin2012}). The addition of extra features such as S\'ersic index undoubtedly helps provide a more accurate separation of early and late type galaxies (\citealt{Driver2006}; \citealt{Cameron2009b}).

Our objective has been to choose a broad range of parameters that will allow us to successfully morphologically classify galaxies with minimal failures. We have been careful to select astrophysically meaningful parameters that denote different aspects of the physicality of a galaxy.  As listed in Table \ref{features}, we have parameters that are known to directly trace galaxy morphology (S\'ersic index, stellar mass, colour), parameters that trace galaxy morphology indirectly (mass-to-light ratio) and parameters that are based on galaxy structure (Kron radius, ellipticity, half-light radius and absolute magnitude). We have attempted to remove the effects of redshift on all the chosen parameters. We also note that in this work, we haven't accounted for the errors in the chosen set of parameters.

The total stellar mass, mass-to-light ratio, absolute magnitude, $g-i$ and $u-r$ colours are taken from the table StellarMassesv18 in the GAMA Data Management Unit (DMU) Stellar Masses \citep{taylor2011}. Total stellar masses have  been derived using Stellar Population Synthesis (SPS) modelling using Bruzual and Charlot models \citep{BC2003} assuming a Chabrier initial mass function \citep{chabrier2005}. SDSS and VISTA-VIKING photometry have been used for this calculation (roughly equivalent to restframe $u - Y$). The mass-to-light ratio has been calculated using the SDSS restframe i-band. The $g-i$ and $u-r$ colours are restframe colours using AB photometry that has been k-corrected to redshift $z=0$ calculated from the Spectral Energy Distribution fit. Together, these colours provide a wide wavelength baseline. Absolute magnitude has been calculated using the restframe r-band from the best SPS SED fit.

Ellipticity, S\'ersic index and half-light radius have been taken from the table SersicCatSDSSv09 in the DMU S\'ersic Photometry \citep{kelvin2012}. These are based on 2-D single S\'ersic function fits to SDSS r-band images.

We obtained Kron radii in arcseconds by multiplying the Kron radius with the angular sizes in semi-major and minor axes and the angular resolution of the main GAMA imaging dataset (0.339"/pixel). These values were converted into kpcs using flow-corrected spectroscopic redshifts from the catalogue DistancesFramesv14 \citep{Baldry2012}.
 
We use morphology for training purposes and to test the robustness of our algorithms. We also note that our parent sample (\citealt{Kelvin2014a};  \citealt{Moffett2016}) is magnitude limited ($M_r <-17.4$ mag) and we do not expect it to be overly sensitive to dwarf galaxy populations. The complete list of parameters that we have used for training and testing are given in Table \ref{features}.

\subsection{Principal Component Analysis} 
\label{pcasec}

\begin{figure} 
\includegraphics[scale=0.5]{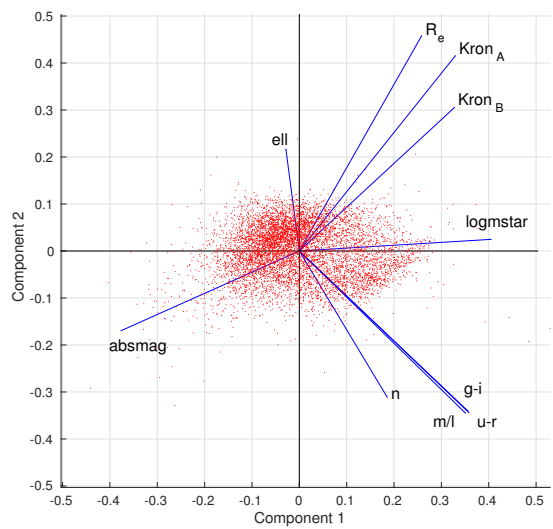}
\caption{Results of Principal Component Analysis (PCA) performed on the selected parameters to determine their impacts on the classification process. The component labels correspond to the parameters given in Table \ref{features} in the following manner: ell = ellipticity; $R_e$ = half-light radius in kpc; $Kron_A$ = Kron radius in kpc (major axis); $Kron_B$ = Kron radius in kpc (minor axis); logmstar = stellar mass; g-i = $g - i$ colour; u-r = $u - r$ colour; $m/l$ = mass-to-light ratio; n = S\'ersic index; absmag = absolute magnitude. Please see Table \ref{features} for more details. The analysis was performed using the MATLAB function {\tt pca}.}
\label{prin}
\end{figure}

We perform Principal Component Analysis (PCA, \citealt{Pearson1901}) on the parameters that we have chosen from the GAMA catalogues (see Section \ref{s_param}, Table \ref{features}). PCA is one of the methods by which parameters are generally chosen for functions such as classification. In our case, we had already defined the criterion for choice of parameters as their distance independence or the possibility of removal of their distance dependence. Therefore, our PCA is a secondary method, to see statistically, the impact each parameter has on the classification process. It was done using the MATLAB function {\tt pca}. Approximately $86\%$ of the variability in our parameters is contained in Components $1-3$ of PCA. For visualisation convenience, we have plotted the first two components in Figure \ref{prin}.

Of the two plotted components, Component 1 contains $\sim57\%$ of the variance of the parameters and Component 2 contains $\sim17\%$. Both stellar mass (logmstar) and absolute magnitude (absmag) have a significant impact on Component 1, but a smaller contribution towards Component 2. The parameters $g-i$ (g-i) and $u-r$ (u-r) colours and mass-to-light ratio (m/l) have very similar contributions to both the components, and are therefore redundant to a great extent. 

Of the other parameters that we have chosen, S\'ersic index (n), Kron radii ($Kron_A$ and $Kron_B$) and half-light radius ($R_e$) seem to have significant contributions toward both Components 1 \& 2, thereby representing sizeable variability in the data set. Ellipticity (ell) seems to be the one with the least variance among our parameters. A detailed analysis of how much each parameter affects the classification process is given in Section \ref{ressec}.

\begin{figure*}
\includegraphics[width=\textwidth]{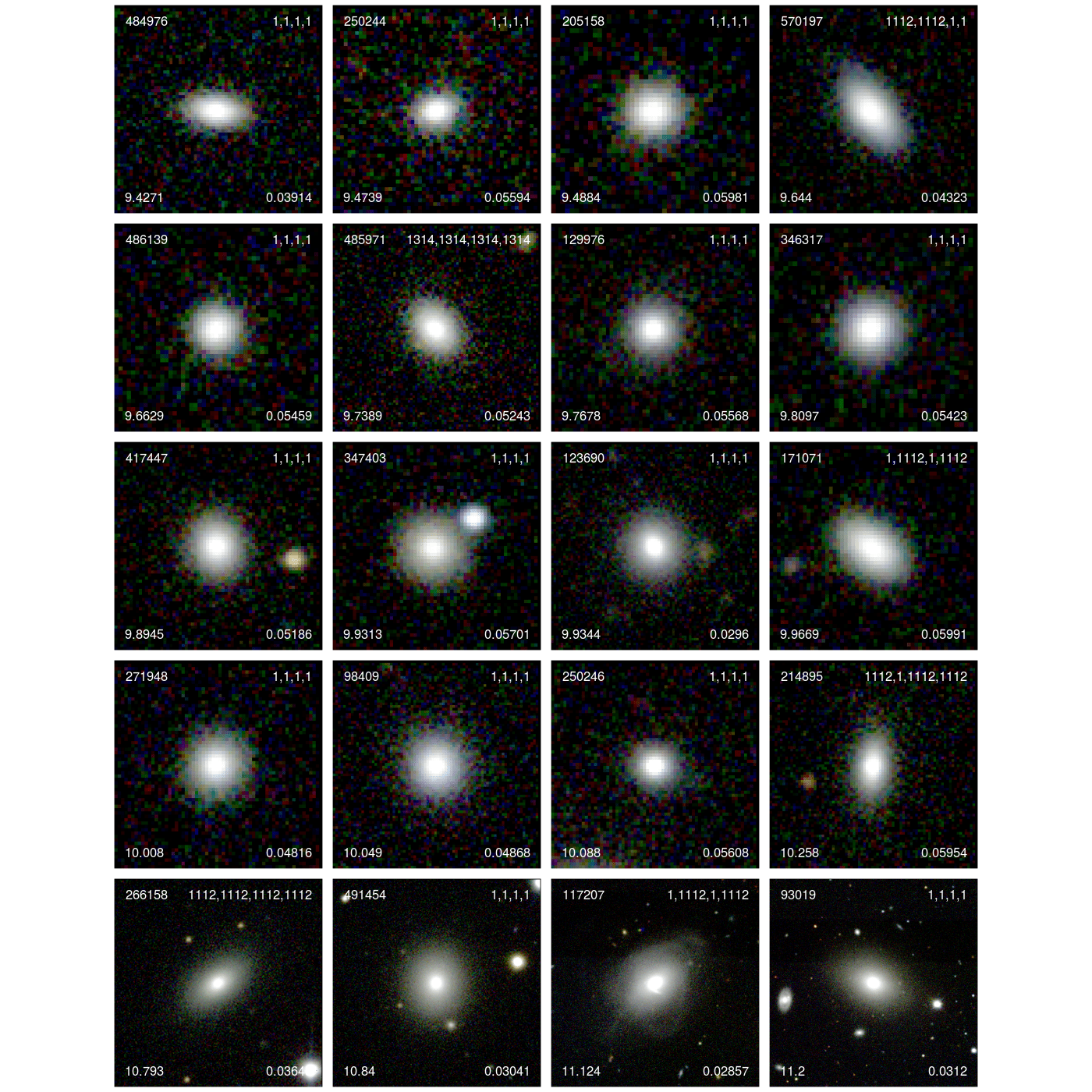}
\caption{A sample of galaxies classified as elliptical (type 1, E) in the GAMA visual morphology catalogue. Postage stamps are log scaled, span an area of 3 $\times$ Kron radius of each galaxy, and are ordered from top-left to bottom-right by increasing stellar mass. Overlaid on each galaxy image are: (top left) the GAMA CATAID of the galaxy; (top right) the numeric Hubble type codes indicating the predicted classification as determined by the SVM, CT, CTRF and NN classifiers, respectively; (bottom left) the total stellar mass in units of $\log_{10}(M_\odot)$, and; (bottom right) the flow corrected spectroscopic redshift of the galaxy. The row-wise median physical scales for these galaxies in kiloparsecs are 5.5, 5.4, 7.4, 7.5 and 2.9. }
\label{image1}
\end{figure*}

\begin{figure*}
\label{image2}
\includegraphics[width=\textwidth]{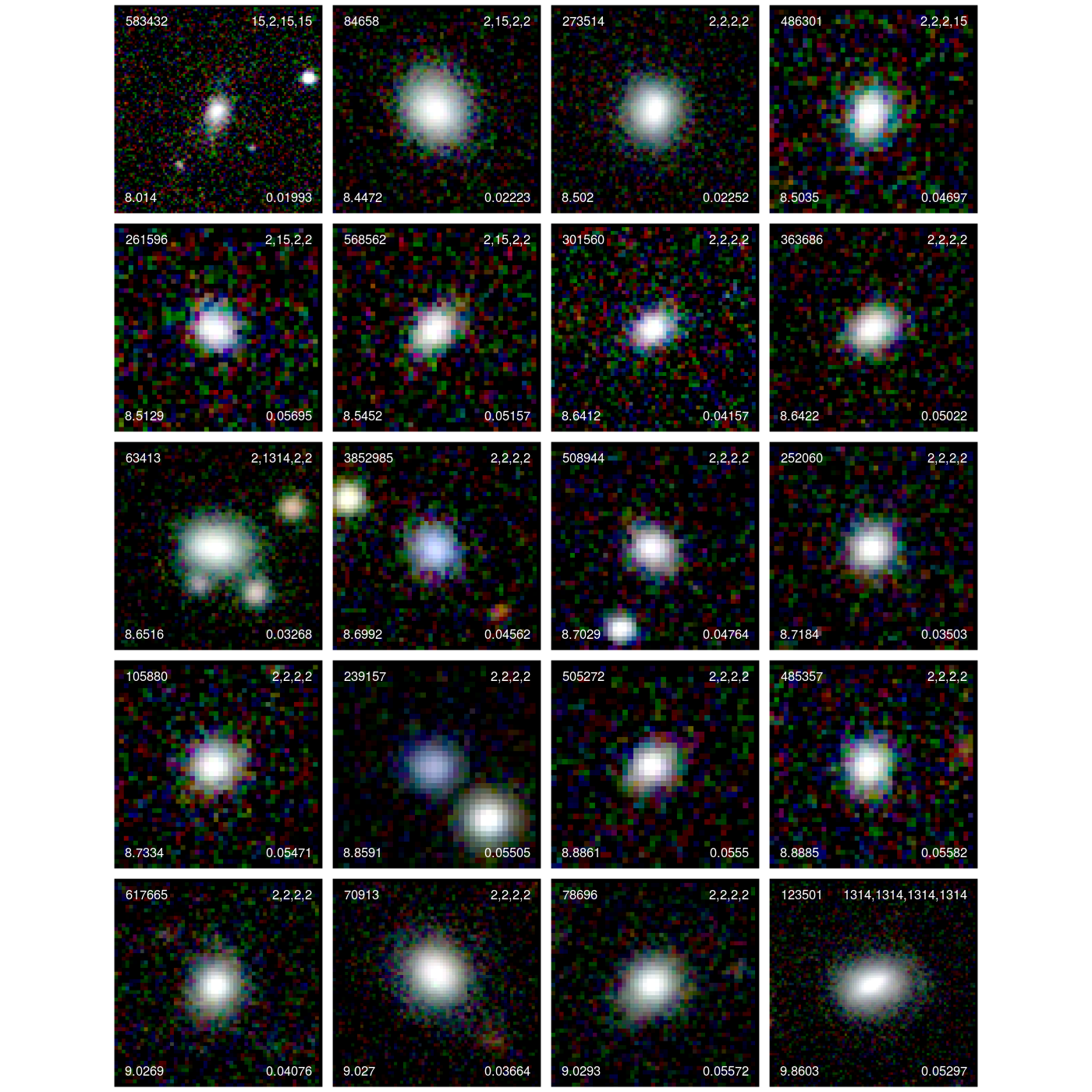}
\caption{As Figure \ref{image1}, but for Little Blue Spheroid (type 2, LBS) galaxies. The row-wise median physical scales for these galaxies in kiloparsecs are 4.6, 5.1, 4.0, 5.0 and 19.0.}
\end{figure*}

\begin{figure*}
\label{image3}
\includegraphics[width=\textwidth]{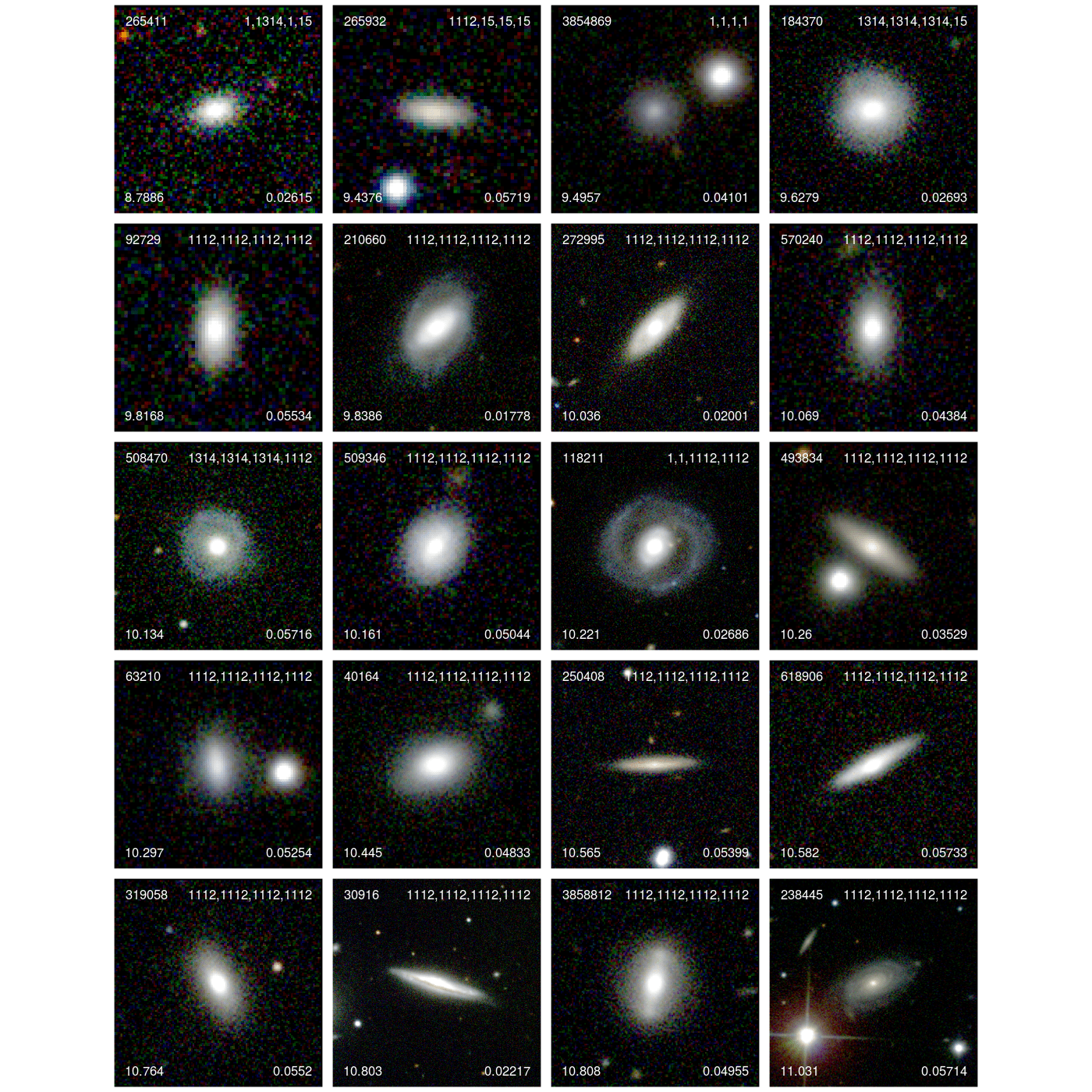}
\caption{As Figure \ref{image1}, but for early-type spiral (type 1112, S0-Sa) galaxies. The row-wise median physical scales for these galaxies in kiloparsecs are 15.9, 24.3, 17.4, 13.5 and 11.8.}
\end{figure*}

\begin{figure*}
\label{image4}
\includegraphics[width=\textwidth]{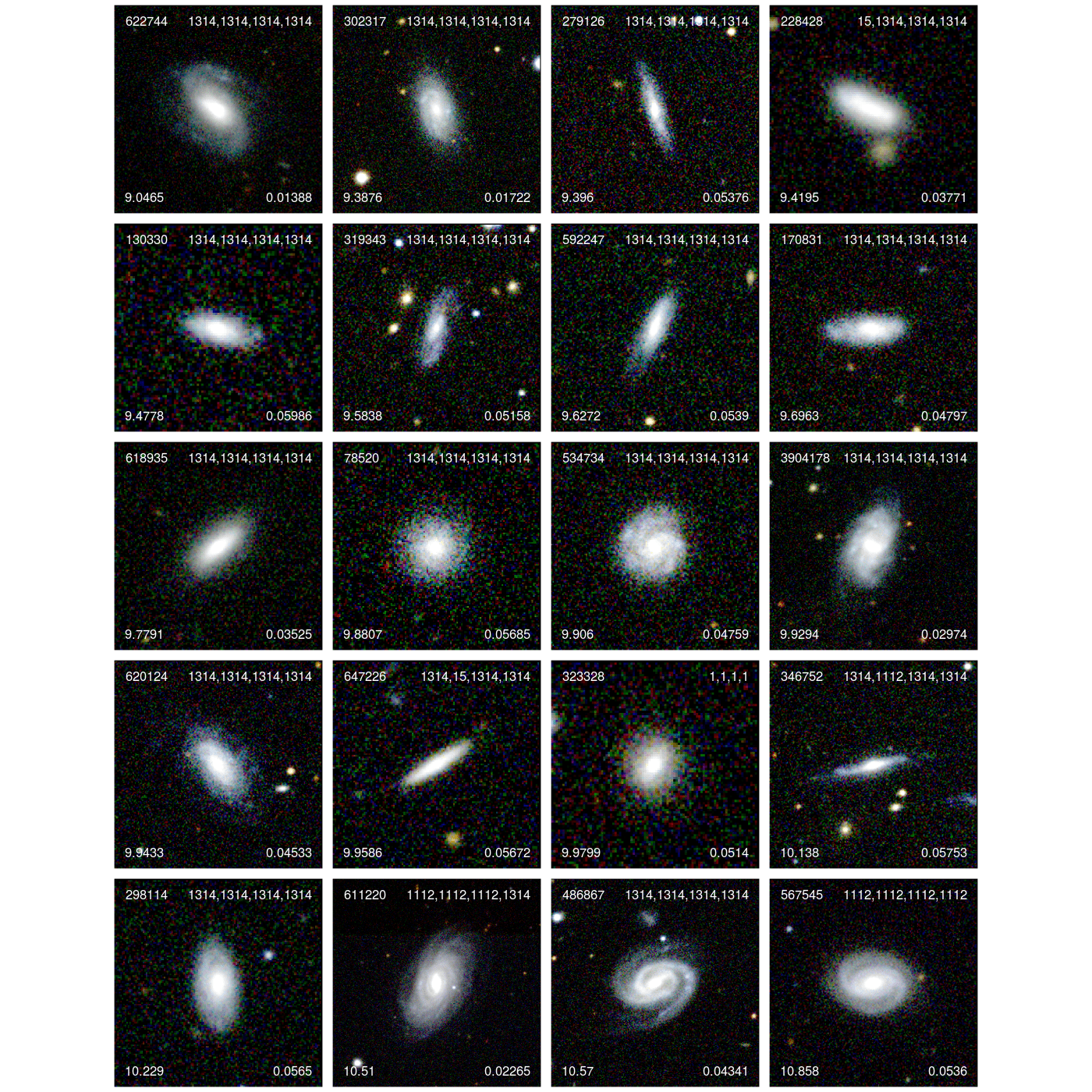}
\caption{As Figure \ref{image1}, but for intermediate-type spiral (type 1314, Sab-Scd) galaxies. The row-wise median physical scales for these galaxies in kiloparsecs are 12.5, 21.8, 12.4, 17.5 and 18.1.}
\end{figure*}

\begin{figure*}
\includegraphics[width=\textwidth]{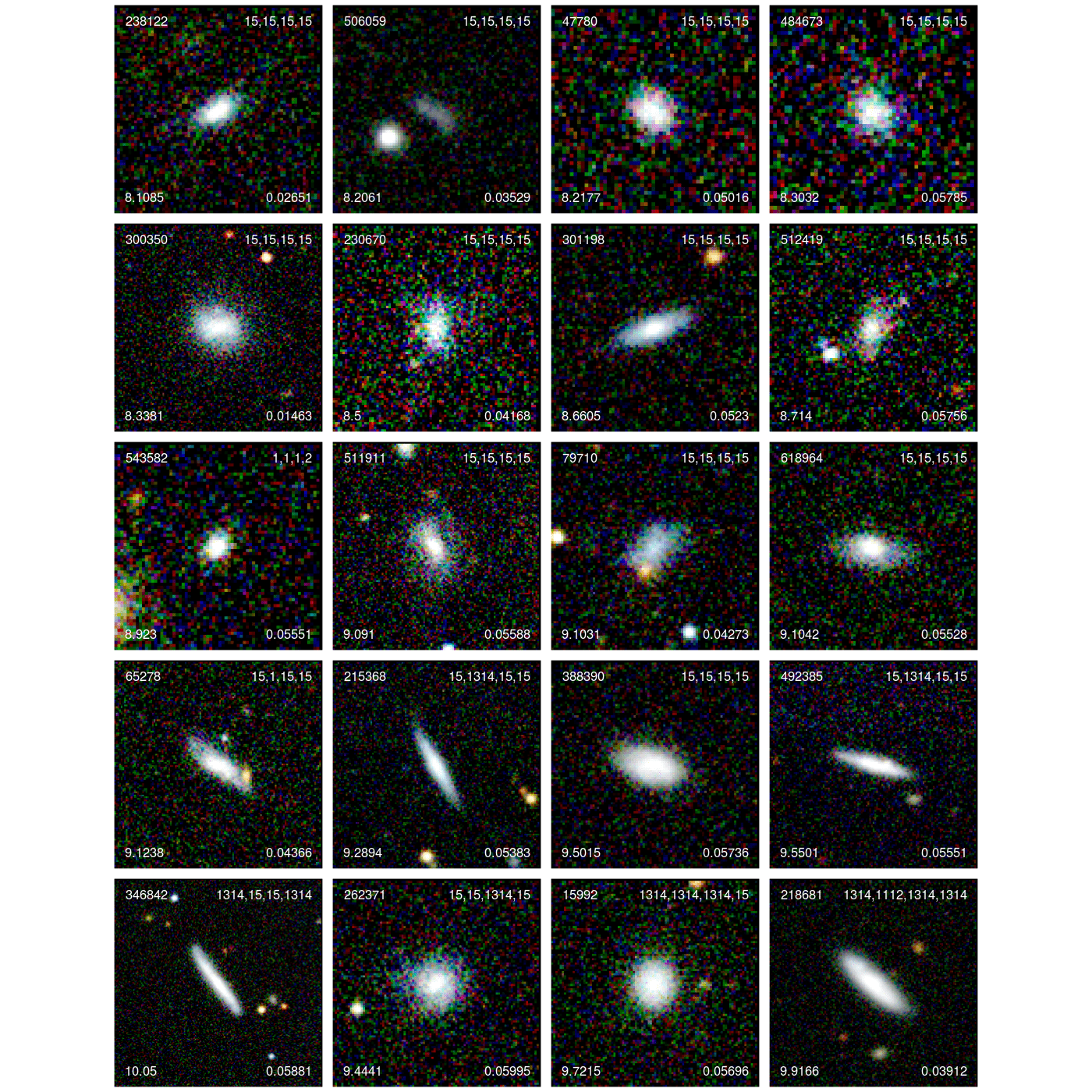}
\caption{As Figure \ref{image1}, but for late-type spiral \& irregular (type 15, Sd-Irr) galaxies. The row-wise median physical scales for these galaxies in kiloparsecs are 28.9, 9.6, 8.1, 12.4 and 17.0.}
\label{image5}
\end{figure*}

\subsection{Data preprocessing}

\label{s_dprep}

Classes $12$ and $14$ are the barred counterparts of classes $11$ and $13$. Their numbers are low in our sample, at $80$ and $195$ respectively. A potential reason for this, as noted in \citet{Kelvin2014a} is that there were noticeable disagreements among the classifiers about the presence of bars in these systems. Another reason could be that, for edge-on systems, it is impossible to verify the presence of bars and therefore they would be classified as unbarred. Due to the relatively low numbers of galaxy systems hosting bars in our sample, we opt to merge the barred classes with their unbarred counterparts. We merge the classes $11$ and $12$ (S0-Sa \& SB0-SBa) to form a new class $1112$. Likewise, we merge classes $13$ and $14$ (Sab-Scd \& SBab-SBcd) to form a new class $1314$. This simplifies the classification problem, albeit marginally. The machine learning classifier that we formulate concentrates on predicting
the GAMA Hubble Types $1$, $2$, $1112$, $1314$ and $15$. Figures \ref{image1}-\ref{image5} show examples of each galaxy type from our final sample. They are created using SDSS g, r, and i band imaging by the GAMA Panchromatic Swarp Imager (PSI) tool\footnote{http://gama-psi.icrar.org/psi.php}. Each image spans a diameter equivalent to 3 $\times$ Kron radius of the galaxy in arcseconds, and is log scaled. 

To construct and evaluate classifiers using statistical learning methods, the data sample is randomly split into training and test sets. The training set is used for constructing classifiers, containing
$80\%$ of the data sample. The test set
is used for the evaluation of the classifiers' prediction abilities, containing the remaining $20\%$ of galaxies. In our case the training and test sets contain $6022$ and $1506$ galaxies respectively. We  consistently use the  same training and test sets for
all considered statistical learning methods described in Section~\ref{s_meths}. The data are normalised before training, i.e. we centre each parameter at its mean value, and scale it to have unit
standard deviation. The distribution of Hubble types for the full data sample, training and test subsets are presented in Table~\ref{gamatypes}.

\section{Methods}
\label{s_meths}
In this section, we outline the galaxy classification problem in the context of statistical learning. We also describe the methods that we apply to solve this classification problem.

\subsection{The classification problem}
\label{classprob}

We consider the parameters of a galaxy to be components of a multidimensional vector
$\xv = \kl{ x_1,x_2,\dots,x_p }^\top\in\R^p$,  where $\kl{ \cdot }^\top$
denotes the transpose of  a vector or matrix. 
Thus, $\xv$ is a  $p\times 1$ column vector. 
In our case $p=10$, and we use the parameters described in Table~\ref{features}.

In the context of statistical learning, the vector space $\R^p$ is often called {\it feature space},
the elements 
$\xv\in\R^p$ are called {\it feature vectors}, and the components $x_i$ of the feature vectors are called {\it features}. The feature vector $\xv$ belongs to one of the $T$ classes.
For convenience, we label the classes as $1,2,\dots,T$.
In our case $T=5$, and the classes correspond to the considered Hubble Types (HT) as $\Set{1,2,1112,1314,15}\widehat{=} \Set{1,2,3,4,5}$.
Let $y\in\Set{ 1,2,\dots,T }$ denote the class label of $\xv$.

Suppose that there is an {\it ideal classifier} $f^*\colon \xv\mapsto y$
that for each feature vector $\xv$ assigns its true classification $y$.
A statistical learning method aims to construct a classifier
$f\colon \xv\mapsto y$ that approximates $f^*$. For this purpose, statistical learning methods use
observational data of the pairs $\kl{  \xv_i, y_i  }$ that contain feature vectors $\xv_i$ for which the corresponding
class $y_i$ is known. A set made up of such pairs $\kl{  \xv_i, y_i  }$ is called the {\it training set}, and we denote it
as $ \zv = \Set{  \kl{\xv_i,y_i},\; i=1,2,\dots,N  }$.

Every statistical learning method consists of a family of classifiers $f$ that depends on certain parameters. 
Using a learning procedure, a particular classifier is chosen from this family
based on the classifier's behaviour on the training data set. 
The selection is typically done such that the classification is well predicted on the training set,
i.e. $f\kl{ \xv_i }\approx y_i$, so as to give low training errors.
The quality of the classifier is then evaluated on the test set, where the classification is known. The data of the test set
is not used for constructing the classifier. Thus, the performance of the classifier on the test set can be seen as an estimation
of its performance on sets with unknown classification.

The methods that we consider here for classifying galaxies are: Support Vector Machines (SVM), Classification Trees (CT), Classification Trees with Random Forest (CTRF) and Neural Networks (NN). We have used the realisation of these methods in MATLAB~R2014b. The outputs provided by the algorithms that we have formulated are multi-class labels, denoting which galaxy type the algorithms deem the galaxy to be of. They are described in detail in the following subsections.

\subsection{Support Vector Machines (SVM)}

The SVM method was originally designed for binary
classification~(\citealt{CriSha00}; \citealt[Chapter~12]{HasTibFri09}). In this method,
for each feature vector $\xv$ there is a 
class label $z \in \Set{-1,1} $.
Therefore for each $\xv_i$ in the training set, the corresponding class is $z_i$. The details of the structure and definitions of the SVM classifier that we employ are given in Appendix \ref{svmapp}.

We use the MATLAB function {\tt svmtrain} for constructing SVM classifiers. For computing the result $f\kl{ \xv }$
of the SVM classifier $f$, function {\tt svmclassify} has been used.

\begin{figure}
\begin{center}

\parbox[t]{8cm}{
\unitlength=1cm
\begin{picture}(8,5)(0,0)
\put(0,0){\includegraphics[width=8cm]{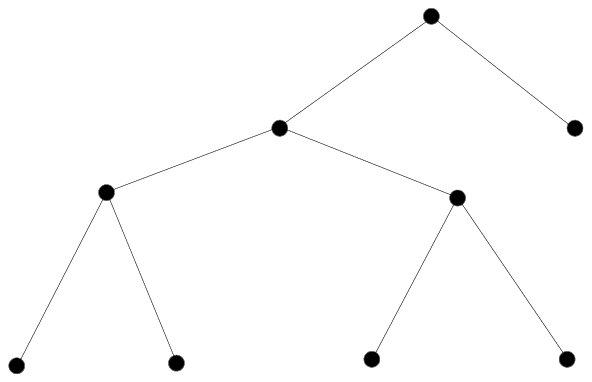}}

\put(1.0,0.3){1}
\put(2.7,0.3){2}
\put(4.4,0.3){1112}
\put(6.4,0.3){1314}
\put(6.6,2.7){15}

\put(1.8,2.8){ $C_{\tm{E}}$ }
\put(5.4,2.7){ $C_{\tm{a}}$ }
\put(3.4,3.4){ $C_{\tm{sp}}$ }
\put(5.0,4.5){ $C_{\tm{15}}$ }

\end{picture}}
\caption{
The binary classification tree determined for the SVM method. The classifier $C_{\tm{15}}$ classifies a galaxy as HT 15 or not. Then, $C_{\tm{sp}}$ classifies into spirals and not spirals. Further classification is done by $C_{\tm{E}}$  into HT 1 (E) or HT 2 (LBS).
$C_{\tm{a}}$ splits the output of the $C_{\tm{sp}}$ classifier into HTs 1112 and 1314. All the binary classifiers in this tree structure are constructed with the SVM method. }
\label{FM_SVMtr}
\end{center}
\end{figure}

In order to use SVM for multi-class classification, the  multi-class 
classification problem is reduced into a series of binary classification problems. 
For this purpose, we consider a tree structure approach \citep{Cam01}. We propose a tree formed by the binary classifiers $C_{\tm{15}}$, $C_{\tm{sp}}$, $C_{\tm{E}}$, $C_{\tm{a}}$  as depicted in
Figure~\ref{FM_SVMtr}. This tree structure is inspired by the distribution of Hubble types in our dataset represented in Table~\ref{gamatypes}. Here, $C_{\tm{15}}$ is the binary classifier that classifies a galaxy as HT 15 or not. $C_{\tm{sp}}$ then classifies into spirals and not spirals. Further classification is done by $C_{\tm{E}}$  into HT 1 (E) or HT 2 (LBS). $C_{\tm{a}}$ splits the output of the  $C_{\tm{sp}}$ binary classifier into HTs 1112 and 1314. All the binary classifiers in this tree structure are constructed with the SVM method. At each binary classifier, the data is split by roughly $50\%$.

\subsection{Classification Trees with hyper-rectangular partitions (CT)}

\begin{figure}
\begin{center}

\parbox[t]{8cm}{
\unitlength=1cm
\begin{picture}(8,7.2)(0,0)
\put(0,0){\includegraphics[width=8cm]{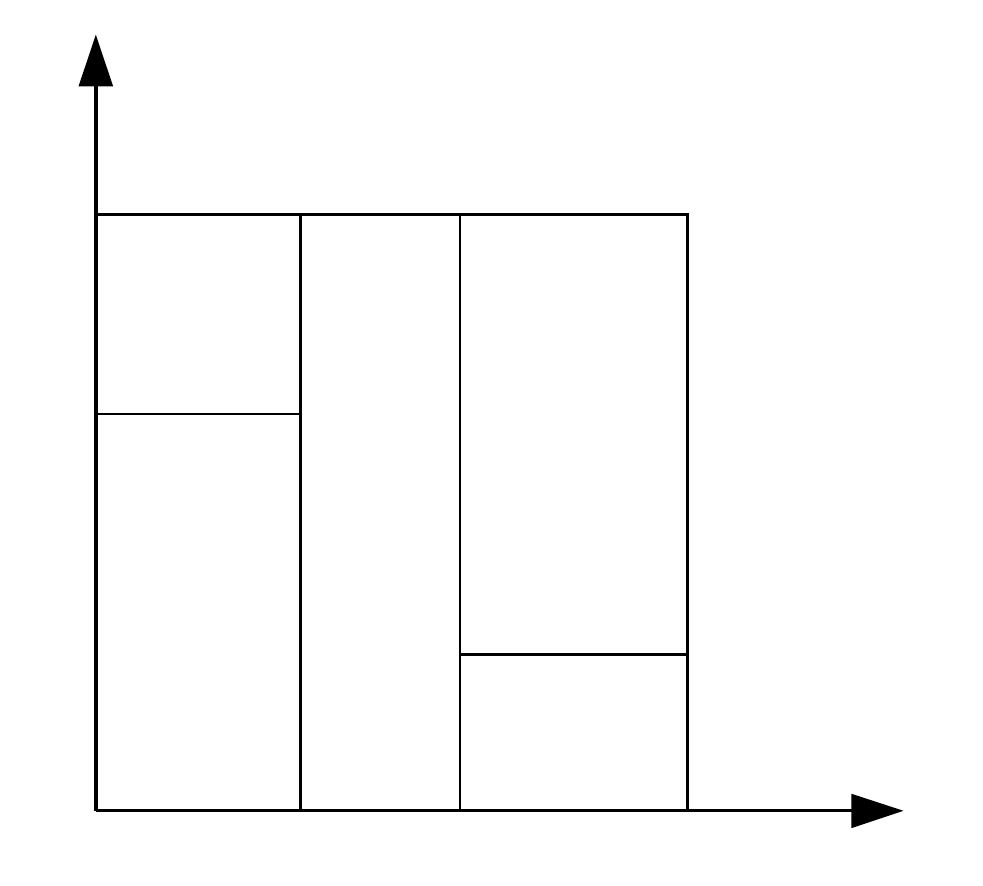}}


\put(6.8,0.0){ $x_1$ }
\put(0.0,6.5){ $x_2$ }
\put(0.5,0.2){ $0$ }
\put(0.3,5.4){ $1$ }
\put(5.4,0.2){ $1$ }

\put(2.2,0.2){ $s_1$ }
\put(3.5,0.2){ $s_3$ }
\put(0.2,3.8){ $s_2$ }
\put(5.7,1.8){ $s_4$ }

\put(1.4,2.0){ $R_1$ }
\put(1.4,4.6){ $R_2$ }
\put(2.9,3.0){ $R_3$ }
\put(4.4,1.2){ $R_4$ }
\put(4.4,3.5){ $R_5$ }

\end{picture}}

\caption{Illustrative example CT method using hyper-rectangular partitions. This unit square is successively split ($s_1 - s_4$) into five nodes R using the two features $x_1$ and $x_2$.
\label{CTpart2}}
\end{center}
\end{figure}
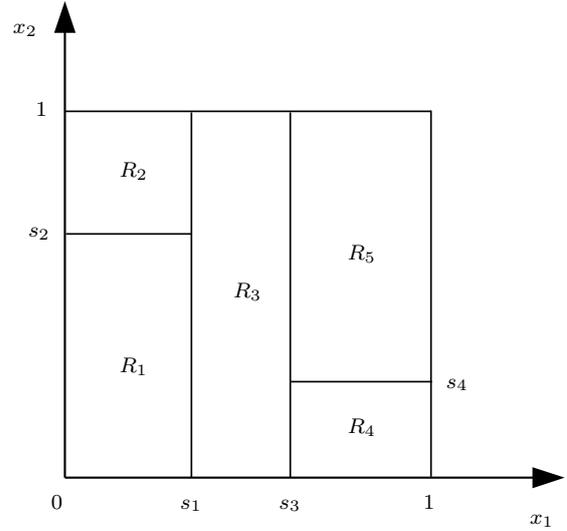

In the CT method the feature space is partitioned into a set of hyper-rectangular regions $R_m$
(\citealt{BreFOS84}, \citealt[Chapter~9]{HasTibFri09}). An example of such a partition is presented in
Figure~\ref{CTpart2}.

The goal of this method is to make the partitions such that each region $R_m$ contains training feature vectors
that belong only to one class, say $k_m\in\Set{1,2,\dots,T}$, or at least the majority of the training feature vectors in $R_m$ is from
one class $k_m$. Then, for each feature vector $\xv$, the CT classifier identifies a region $R_m$ that contains
$\xv$, and then assigns $k_m$ as the predicted class for $\xv$. The method is discussed in detail in Appendix \ref{ctapp}.

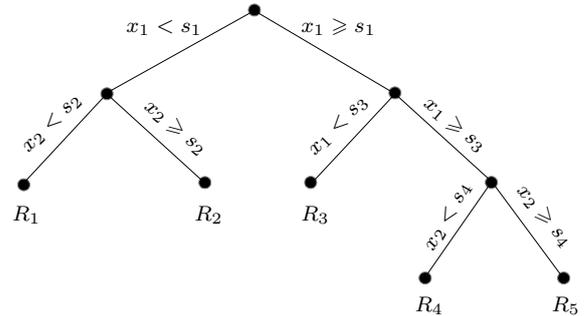
\begin{figure}
\begin{center}

\parbox[t]{8cm}{
\unitlength=1cm
\begin{picture}(8,5.5)(0,0)
\put(0,0){\includegraphics[width=8cm]{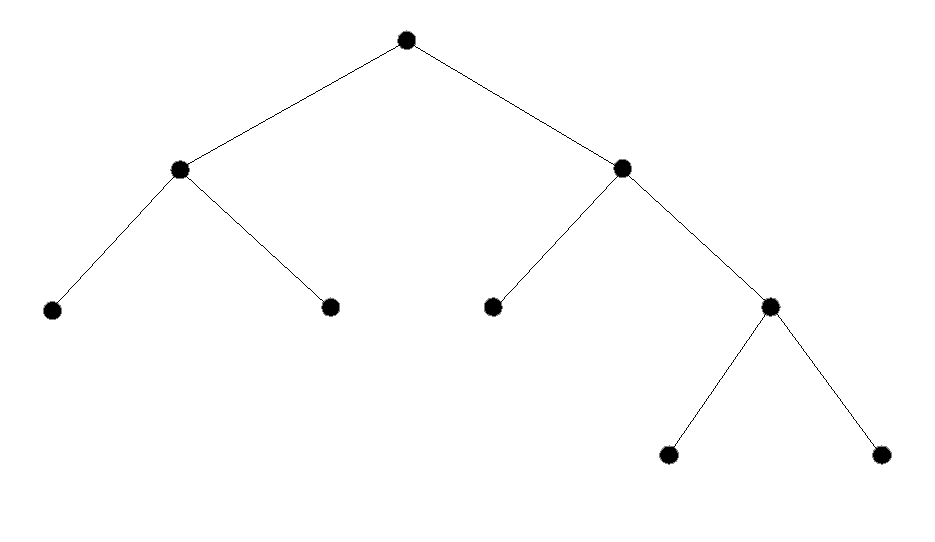}}


\put(0.2,1.5){ $R_1$ }
\put(2.6,1.5){ $R_2$ }
\put(4.0,1.5){ $R_3$ }
\put(5.5,0.3){ $R_4$ }
\put(7.3,0.3){ $R_5$ }

\put(1.7,4.0){ $x_1 < s_1$ }
\put(4.0,4.0){ $x_1 \geq s_1$ }

\put(4.0,2.3){ \rotatebox{45}{ $x_1 < s_3$ }}
\put(5.5,3.1){ \rotatebox{-40}{ $x_1 \geq s_3$ }}

\put(5.6,1.1){ \rotatebox{60}{$x_2 < s_4$} }
\put(6.8,1.9){ \rotatebox{-50}{$x_2 \geq s_4$} }

\put(0.3,2.4){ \rotatebox{45} {$x_2 < s_2$} }
\put(1.9,3.0){ \rotatebox{-40} {$x_2 \geq s_2$} }

\end{picture}}

\caption{A binary classification tree determined for the CT method as applied to the example unit square shown in Figure~\ref{CTpart2}.
\label{CTtree}}
\end{center}
\end{figure}

The CT partitioning can also be represented by a binary tree, i.e., the partition presented in Figure~\ref{CTpart2}
can be represented by the tree in Figure~\ref{CTtree}. The top node of the tree, which is called root, represents the complete
feature space. Feature vectors that satisfy the condition $x_1<s_1$ are assigned to the next lower node on the left,
while the other feature vectors are assigned to the next lower node on the right, and so on.
The nodes at the bottom of the tree, which are called terminal nodes or leaves, correspond to the regions of the final
partition of the feature space: $R_1$, $R_2$, $\dots,$$R_5$.

The node splitting is recursively repeated for the new nodes. The node is not split if any of the following conditions is satisfied:

\begin{itemize}
\item The node is pure.

\item The node contains less than a certain number (standard value adopted here is 10) of training feature vectors.

\item Any node splitting gives new nodes that contain less or equal to a certain number (standard value adopted here is 0) of training feature vectors.

\item If a certain number of nodes (the default value for the MATLAB function that generates the node splitting is N-1) are created.

\end{itemize}

For our work, we constructed the CT classifier using the MATLAB function {\tt fitctree} and the function {\tt predict} was used for computing the result of the CT classifier.
In the constructed CT classifier for our dataset a full description of the derived nodal splits becomes increasingly complex beyond the first leaf. Therefore we describe the splits which were determined up to and including the first leaf only. The splitting feature in the top node (i.e., at the root of the constructed tree) is $x_1$ which corresponds to the
stellar mass of a galaxy. The split point for this feature was determined to be $\log M = 9.276$. The next leaf node (in the regime $x_1 < 9.276$) has the splitting
feature $x_6$, which is the half-light radius, with the split point determined to be $\log R_e= 0.0514$. The alternative node (i.e., the galaxies in the regime $\log M \geq 9.276$) has the
splitting feature $x_8$, which is $u-r$ colour, with the split point $ u-r = 1.842$.

The structure of the classifier in the CT method is quite simple.
Notably, no arithmetic operation is used
for estimating the class of the feature vector $\xv$. Only a comparison between numbers is used.
Therefore, the evaluation of the result of the CT classifiers is very fast, which is a distinct advantage of this method.

However, CT classifiers are known to have the following drawback.
$f\kl{ \xv_i }$ can be in a good agreement with $y_i$, but outside the training set, the predictive performance of the CT classifier
may be rather poor. This phenomenon is called {\it overfitting}. To overcome this drawback, the idea of {\it Random Forest}
has been proposed~(\citealt{HasTibFri09}, Chapter~15; \citealt{Bre01}).
This leads to the CTRF method that we explore in the next subsection.


\subsection{Classification Trees with Random Forest (CTRF)}


The essential idea of the CTRF method is to improve the performance of a single CT by averaging over several differently trained CTs. In order to achieve this, a certain number of samples are created by random sampling {\it with replacement} from the training set. The sampling is done using uniform distribution, where each sample is of the same size as the original training set. By using sampling with replacement, any element of the training set can be selected more than once for the same random sample. More details on this process are given in Appendix \ref{ctrfapp}.

Each CT classifier in a Random Forest is trained on a different sample of the training data. Moreover, the use of the modified CT learning algorithm, namely the use of random subsets of the features, ensures the {\it de-correlation} between the constructed CT classifiers. This means that the tree structure of the involved CT classifiers differ from one CT to another. These two properties allow the combination via majority vote of the CTs in the Random Forest to correct the overfitting of each CT classifier. For building our CTRF classifier, we used the MATLAB class {\tt TreeBagger}, and the function {\tt predict}
was used for calculating the outcome of the CTRF classifier.

The choice of the number of samples $B$ in Random Forests can be done by observing the {\it out-of-bag error}.
This error is the mean prediction error on each training example using only the CT classifiers that did not have this example in their
training sample \citep[p.~593]{HasTibFri09}. In our case, we observed that this error stabilises for $B=100$, and therefore, we used this number for our CTRF classifier.


\subsection{Single hidden layer feed forward Neural Networks (NN)}

The last statistical learning method that we consider is Neural Networks (NN)~(\citealt[Chapter~11]{HasTibFri09}). This is a classification method inspired by the central nervous system or biological neural networks of animals.
In comparison to the other mentioned methods, NN constructs classifiers with a more complicated mathematical structure,
and the algorithms for constructing NN classifiers are more complex.
However, a typically good performance of the NN classifiers outside the training sets makes them
very popular.

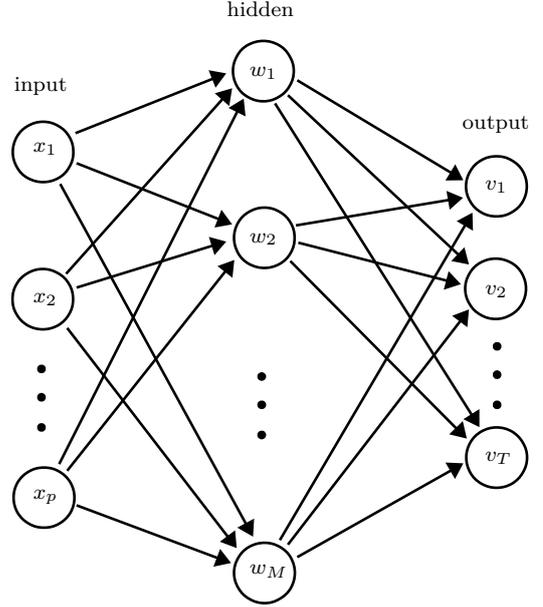
\begin{figure}
\begin{center}

\parbox[t]{8cm}{
\unitlength=1cm
\begin{picture}(8,9)(0,0)
\put(0,0){\includegraphics[width=8cm]{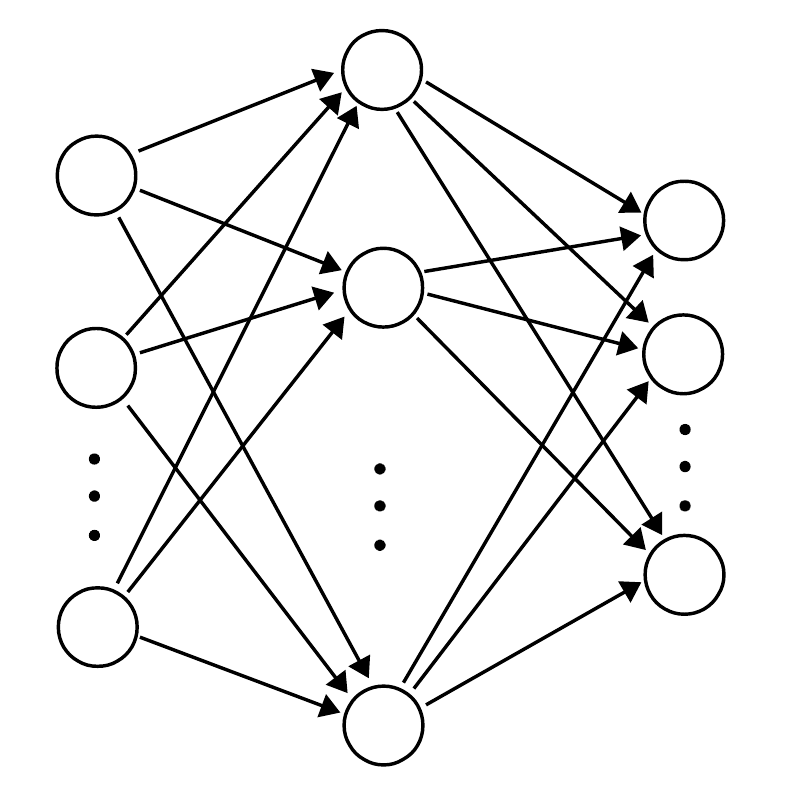}}


\put(0.85,1.8){$x_p$}
\put(0.85,4.4){$x_2$}
\put(0.85,6.4){$x_1$}
\put(0.6,7.2){input}

\put(3.7,0.8){$w_M$}
\put(3.7,5.2){$w_2$}
\put(3.7,7.4){$w_1$}
\put(3.4,8.2){hidden}

\put(6.8,2.3){$v_T$}
\put(6.8,4.5){$v_2$}
\put(6.8,5.9){$v_1$}
\put(6.5,6.7){output}


\end{picture}}

\caption{A network diagram for the single hidden layer feed-forward neural network.
\label{FM_NN}}
\end{center}
\end{figure}

A neural network consists of units that are organized in layers. Typically, a network diagram, such as in
Figure~\ref{FM_NN}, is used to represent a neural network. In this work, we implement the most widely used neural network ensemble called the {\it single hidden layer feed-forward neural network}. It consists
of three layers, the input layer, hidden layer, and output layer.


The units in the input layer correspond to the features $x_i$. The $k$th unit $v_k$ in the output layer
models the probability for the feature vector to belong to class $k$. The units in the hidden layer
$w_m$, $m=1,2,\dots,M$, can be seen as additional features that are derived from the features $x_i$. The structure of the neural network that we have considered is explained in more detail in Appendix \ref{nnapp}.

For defining our NN classifier, we used the MATLAB function {\tt patternnet}. Then, the weights of the NN classifier
were determined using the function {\tt train}, and the evaluation of the result of the classifier was performed. We consider values for the number of units in the hidden layer $M$ in the interval [10,500] and examine the performance of the corresponding NN classifiers on the so-called validation set. For this set, we randomly sample 15\% of the elements in the training set. These elements were not used for training the NN classifiers. We find that the True Prediction Ratio (TPR) for the validation set increases as a function of $M$; however, the relative increase in TPR significantly diminishes as we tend towards larger values of $M$. We therefore adopt $M=500$ as the optimal trade-off between classification accuracy and computational complexity of the NN classifier.






\section{Results}
\label{ressec}

\begin{table*}

\ctw=0.8cm
\ctwa=1.5cm
\caption{True Prediction Ratios (TPRs) in percentages for the classifiers obtained by the methods considered in Section~\ref{s_meths} on the test set are given in panel 1. Panel 2 represents the results of binary classification using CTRF method. The galaxy types E, LBS \& S0-Sa are collectively considered as spheroid-dominated systems and Sab-Scd \& Sd-Irr as disk-dominated systems.}
\resizebox{0.8\textwidth}{!}
{\begin{tabular}{lllllll}
\hline
\\
HT & \ct {0.3cm}{E \\1} &\ct{0.3cm}{LBS\\ 2}& \ct{0.8cm}{S0-Sa\\1112}& \ct{1.1cm}{Sab-Scd\\1314}& \ct{0.8cm}{Sd-Irr\\15} & All\\
\\
\hline
\\
CT & $61.5^{+3.5}_{-3.8}$ & $63.3^{+3.4}_{-3.7}$ & $56.3^{+3.7}_{-3.8}$ & $52.9^{+3.0}_{-3.0}$ & $82.0^{+1.4}_{-1.6}$ & $69.0^{+1.2}_{-1.2}$ \\
\\
CTRF & $70.7^{+3.2}_{-3.7}$ & $75.6^{+2.9}_{-3.5}$ & $63.6^{+3.5}_{-3.8}$ & $56.4^{+2.9}_{-3.0}$ & $88.9^{+1.1}_{-1.3}$ & $76.2^{+1.1}_{-1.1}$ \\
\\
SVM & $70.1^{+3.2}_{-3.7}$ & $76.7^{+2.9}_{-3.4}$ &$63.6^{+3.5}_{-3.8}$ & $53.2^{+3.0}_{-3.0}$ & $89.2^{+1.1}_{-1.3}$ & $75.8^{+1.1}_{-1.1}$\\
\\
NN & $67.2^{+3.4}_{-3.7}$ & $72.2^{+3.1}_{-3.6}$& $62.5^{+3.5}_{-3.8}$ & $57.9^{+2.9}_{-3.0}$ & $89.8^{+1.0}_{-1.3}$ & $76.0^{+1.1}_{-1.1}$\\
\\
\hline

\multicolumn{1}{c}{} & 
    \multicolumn{3}{@{}c@{}}{$\underbrace{\hspace*{\dimexpr18\tabcolsep+2\arrayrulewidth}\hphantom{012}}_{}$} & 
    \multicolumn{2}{@{}c@{}}{$\underbrace{\hspace*{15\tabcolsep}\hphantom{3}}_{}$}\\
    

&\multicolumn{3}{c}{Spheroid-dominated}&\multicolumn{2}{c}{Disk-dominated}&All\\
\\
\hline
\\
CTRF&\multicolumn{3}{c}{$84.9^{+1.4}_{-1.7}$}&\multicolumn{2}{c}{$92.5^{+0.8}_{-0.9}$}&$89.8^{+0.7}_{-0.8}$\\
\\
\hline

\end{tabular}}

\label{T_TPRs}
\end{table*}

\begin{figure*}
\includegraphics[scale=0.6]{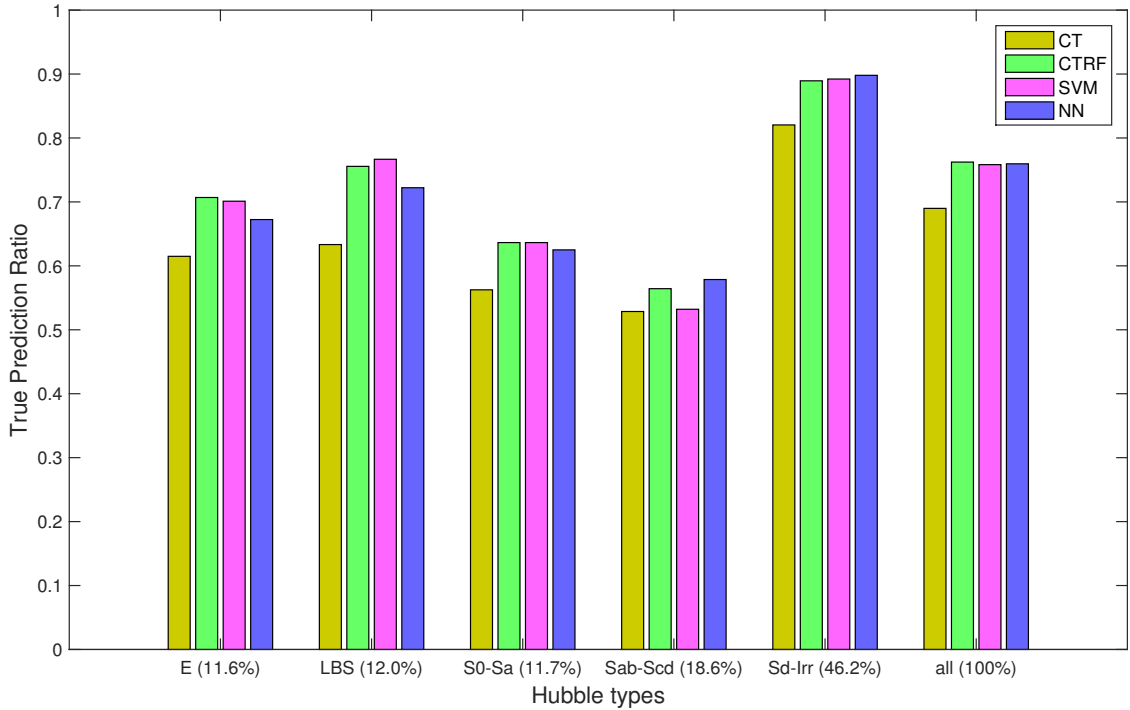}
\caption{Histograms showing the True Prediction Ratios from panel 1 of Table \ref{T_TPRs}. The different Hubble types in our sample are represented on the x-axis and the TPR values for each type as obtained by the four statistical learning algorithms are shown on the y-axis. The percentage of galaxies of a certain type are shown in brackets next to the Hubble type codes. }
\label{result}
\end{figure*}

\begin{figure*}
\includegraphics[scale=0.8]{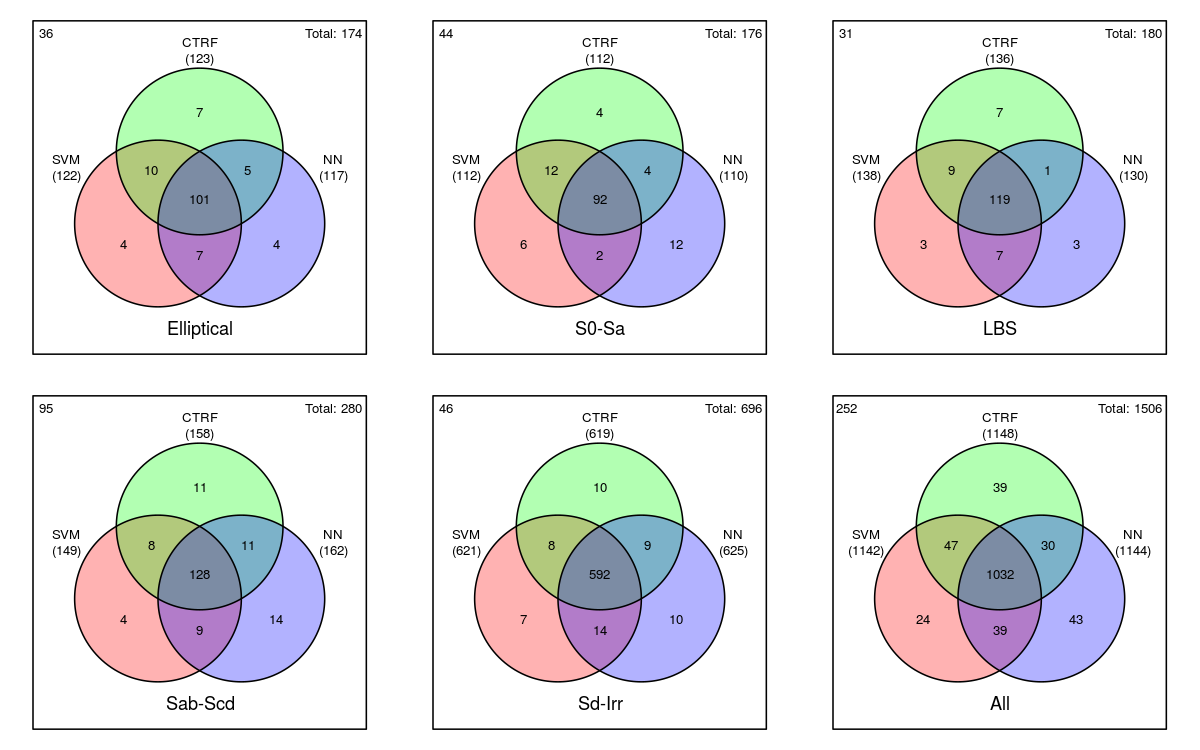}
\caption{Venn diagrams representing the effectiveness of classification by CTRF, SVM and NN methods for each GAMA Hubble type and over all types. The number of objects `correctly' classified by each method is shown in brackets next to the algorithm labels. The number of objects which were not classified `correctly' by any method is shown in the top left corner while the total number of objects is given in the top right corner.}
\label{venn}
\end{figure*}

\begin{figure*}
\includegraphics[width=\textwidth]{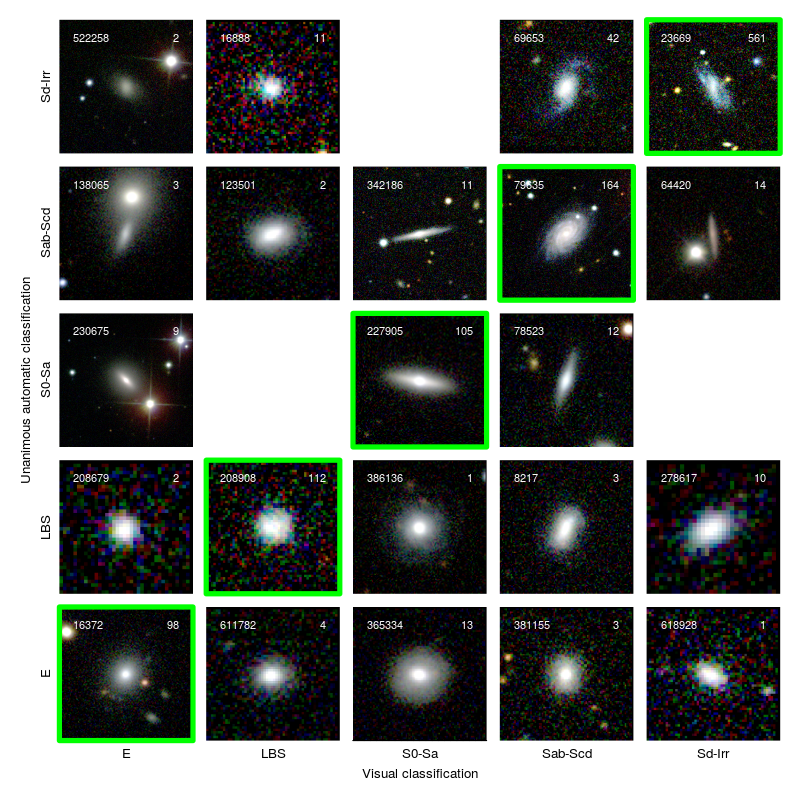}
\caption{Figure illustrating unanimous disagreement. The x-axis represents the visual classification of the objects while the y-axis shows the unanimous automatic classifications. For example, the galaxy in the bottom most row with ID 611782 has been visually classified as LBS while all four algorithms used in this study classify it as type E. The prime diagonal represents objects for which the visual classification and the four algorithms are in agreement (highlighted in green). The number of objects in each bin is noted in the top right corner of each postage stamp. The other blank spaces denote the absence of objects of x-axis type unanimously classified by the four algorithms as the y-axis type. }
\label{undis}
\end{figure*}

\begin{figure*}
\makebox[\textwidth]{\includegraphics[width=\textwidth]{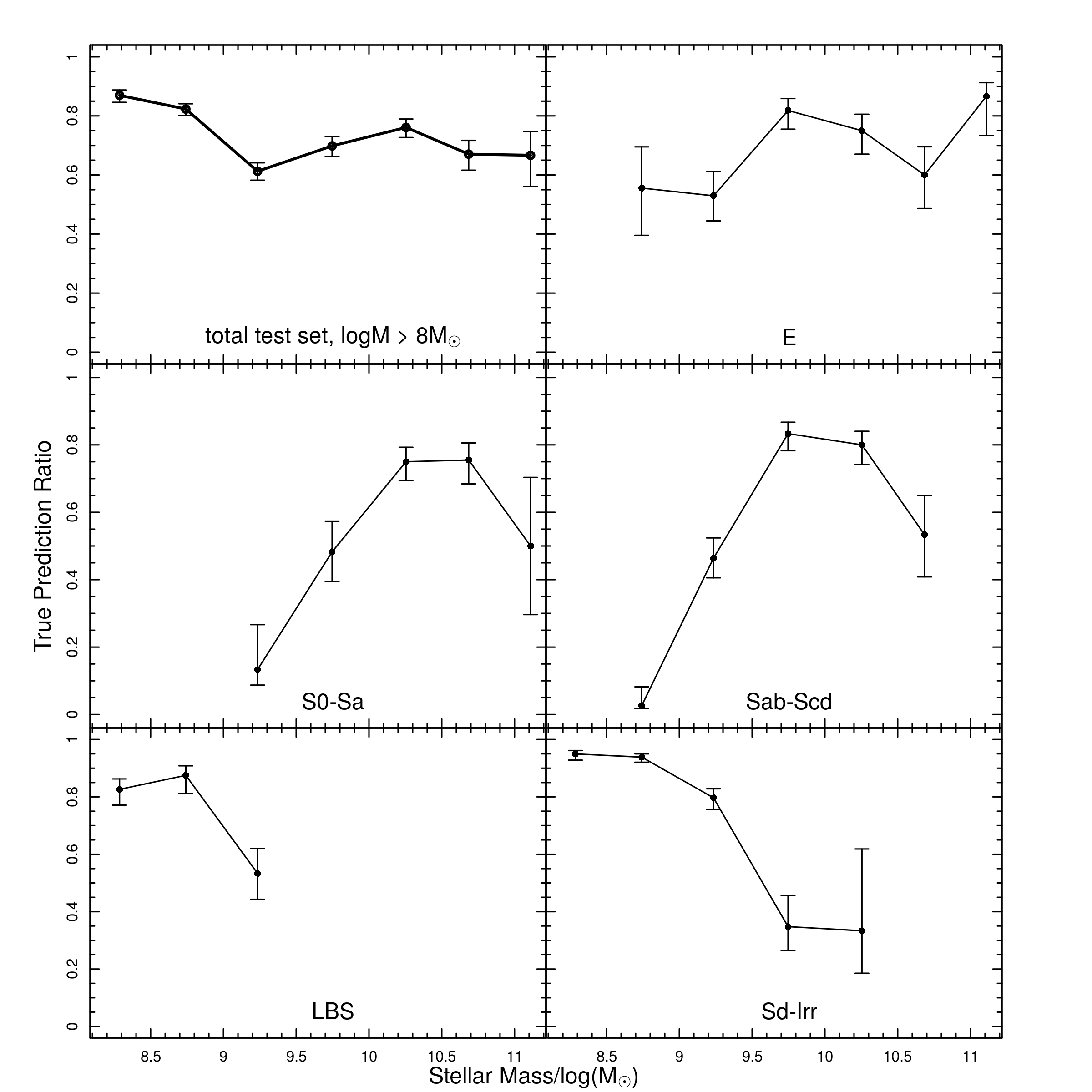}}
\caption{Representation of the TPR as a function of total stellar mass (log) for the method that we recommend, CTRF. The distribution over the total test set is represented in the first panel. The individual contributions of the different GAMA Hubble types are plotted in the subsequent panels as indicated. The lower and upper boundary fractional errors for the data set are calculated by using the {\tt aqbeta} function from the astro library in R \citep{Cameron2011}.}
\label{xbox}
\end{figure*}

\begin{figure*}
\makebox[\textwidth]{\includegraphics[width=\textwidth]{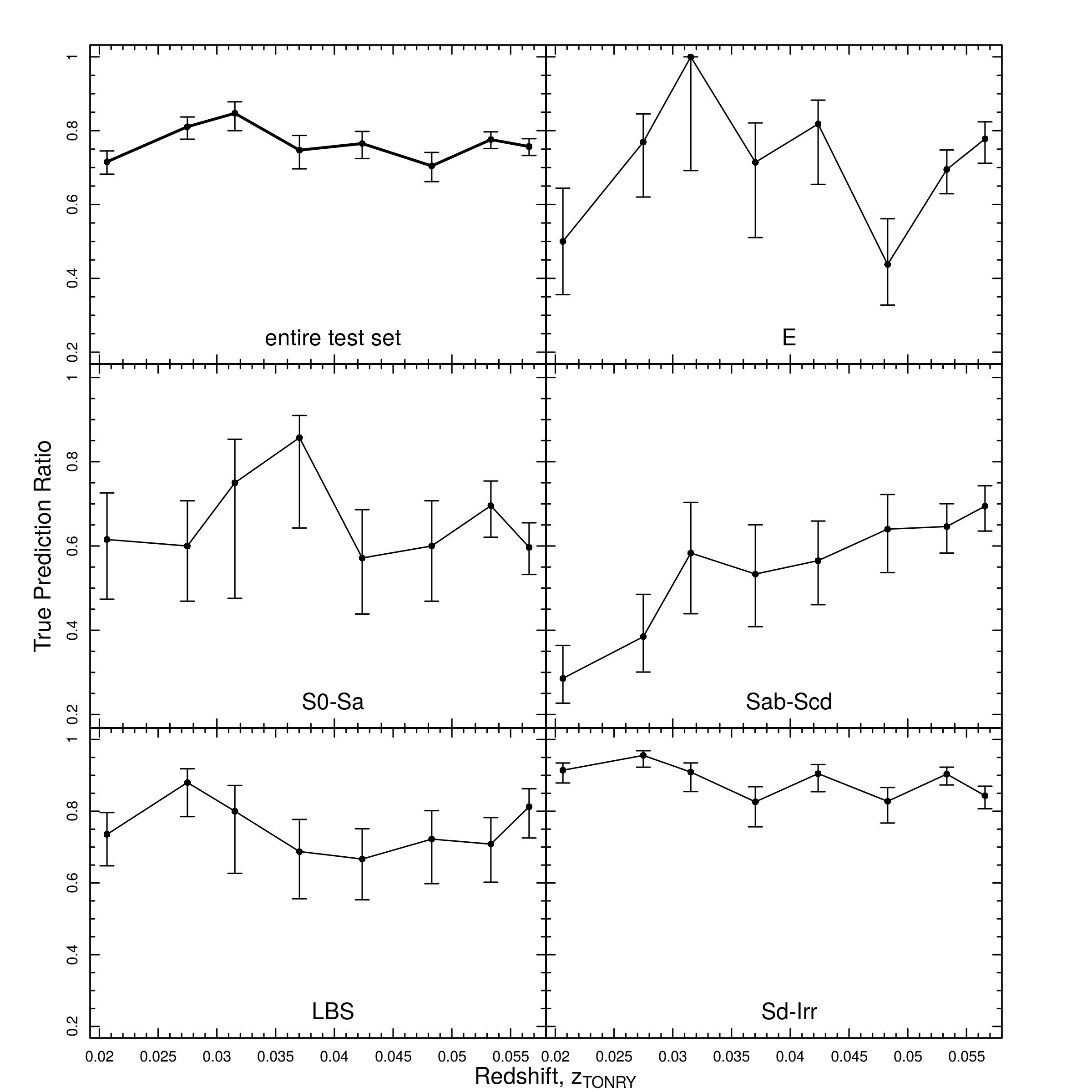}}
\caption{As Figure \ref{xbox}, but as a function of redshift. }
\label{tprz}
\end{figure*}

The CT, CTRF, SVM and NN codes are run using the parameters shown in Table \ref{features}. Figure \ref{result} shows the classification success rate for each morphological type considered in addition to the total sample (`all'). Galaxy populations are arranged along the x-axis, as indicated. Classification success rate is characterised by the parameter True Prediction Ratio (TPR) shown on the y-axis. TPR (y-axis)\footnote{Here onward, this parameter is used interchangeably with accuracy of classification \citep{soklap09}.} represents the quality measure of the classifiers. It is defined as the ratio of the  number of correctly classified galaxies to the total number of galaxies considered. The TPR for the machine learning algorithms CT, CTRF, SVM and NN are represented by the colours yellow, green, pink and blue, respectively, for each morphological type. As can be seen, the morphological type Sd-Irr (Type 15) typically returns the highest success ratio at $\sim90\%$. The morphological type Sab-Scd (Type 1314) returns the lowest average success ratio, typically in the range $\sim55\%$. Potential reasons for this are discussed in detail in Section~\ref{disc}, but principally revolve around the idea that our algorithms in their current configuration may be more suited to classify single component rather than more complex multi-component systems. The overall average success rate across all morphological types is found to be $\sim76\%$, with the notable exception of the CT method (see Table~\ref{T_TPRs}).

Classification errors can be also characterised using a {\it confusion matrix}, $  \kl{  a_{ij}  }_{i,j = 1}^T  $. The entry of this matrix $a_{ij}$ in the $i$-th row and $j$-th column
is the number of galaxies from the class $j$ that are classified as the class $i$ by the classifier.

Note that the above considered quality measure TPR of a classifier for the class $j$ can be calculated
using the confusion matrix $  \kl{  a_{ij}  }_{i,j = 1}^T  $ of this classifier :
$$
  \tm{TPR}_j = \frac{  a_{jj}  }{  \sum_{i=1}^T  a_{ij}  }.
$$
This quality measure is also known under the names {\it true positive rate} or {\it recall}.

The TPR of a classifier for all classes is calculated as
$$
  \tm{TPR}_{\tm{all}} = \frac{   \sum_{j=1}^T  a_{jj}  }{  \sum_{i,j=1}^T  a_{ij}  }.
$$

In addition to the TPR, another useful characteristic of the classifier performance is the {\it  Positive Predictive Value } (PPV) or {\it precision}. It is calculated for the class $j$ using the confusion
matrix $  \kl{  a_{ij}  }_{i,j = 1}^T  $ :
$$
  \tm{PPV}_j = \frac{  a_{jj}  }{  \sum_{i=1}^T  a_{ji}  }.
$$

Another important characteristic is the {\it F-score} of the classifier. For the class $j$ it is defined as the
harmonic mean of $\tm{TPR}_j$ and $\tm{PPV}_j$:
$$
  \tm{F}_j = \frac{   2\cdot \tm{TPR}_j  \cdot  \tm{PPV}_j   }{   \tm{TPR}_j  +  \tm{PPV}_j   }.
$$

The confusion matrices and the mentioned performance characteristics of the considered classifiers are presented
in Tables~\ref{T:SVM}--\ref{T:bCTRF}. The actual classification is given in the columns and the classification predicted by the classifiers in rows. The rows and columns represent the five galaxy types.


%
\begin{table}
\begin{center}

Visual classification
\skpln

\begin{tabular}{cccccccc}
&&&  E&LBS&S0-Sa&Sab-Scd&Sd-Irr\\
&&E &122&		12&		35&		9&		7 \\
&&LBS &13&		138&		3&		12&		30\\
&&S0-Sa &22&		0&		112&		30&		2 \\
&&Sab-Scd &10&		3&		24&		149&		36\\
\rotatebox{90}{\parbox{3mm}{\multirow{3}{*}{ \ct{0.3cm}{SVM\\ classification}}}}&&Sd-Irr &7&		27&		2&		80&		621
\end{tabular}


\skpln
performance characteristics
\skpln

\begin{tabular}{cccccc}
\hline
&E&LBS&S0-Sa&Sab-Scd&Sd-Irr\\
TPR&    70.1&    76.7&    63.6&    53.2&    89.2\\
PPV&    66.0&    70.4&    67.5&    67.1&    84.3\\
F&	   68.0 &   73.4&    65.5&    59.4&    86.7 \\ \hline
\end{tabular}

\end{center}

\caption{
Confusion matrix and performance characteristics for 5 galaxy classes for the SVM classifier.
\label{T:SVM}}
\end{table}
%


%
\begin{table}
\begin{center}

Visual classification
\skpln

\begin{tabular}{cccccccc}
&&&  E&LBS&S0-Sa&Sab-Scd&Sd-Irr\\
&&E &107	&	21&		38&		11&		17 \\
&&LBS &4&		114&		3&		8&		41 \\
&&S0-Sa &37&		3&		99&		34&		9 \\
&&Sab-Scd &15&		9&		31&		148&		58\\
\rotatebox{90}{\parbox{3mm}{\multirow{3}{*}{ \ct{0.3cm}{CT\\ classification}}}}&&Sd-Irr &11&		33&		5&		79&		571
\end{tabular}


\skpln
performance characteristics 
\skpln

\begin{tabular}{cccccc}
\hline
&E&LBS&S0-Sa&Sab-Scd&Sd-Irr\\
TPR&    61.5 &   63.3 &   56.3 &   52.9 &   82.0\\
PPV&    55.2 &   67.1 &   54.4 &   56.7 &   81.7\\
F	 &  58.2 &   65.1 &   55.3 &   54.7 &   81.9
\\ \hline
\end{tabular}

\end{center}

\caption{
As for Table \ref{T:SVM}, but for the CT classifier.
\label{T:CT}}
\end{table}
%


%
\begin{table}
\begin{center}

Visual classification
\skpln

\begin{tabular}{cccccccc}
&&&  E&LBS&S0-Sa&Sab-Scd&Sd-Irr\\
&&E&123&		15&		31&		5&		11\\
&&LBS&8&		136&		4	&	10&		31\\
&&S0-Sa&24&		1	&	112&		25&		2 \\
&&Sab-Scd&8&		2	&	26&		158	&	33\\
\rotatebox{90}{\parbox{3mm}{\multirow{3}{*}{ \ct{0.3cm}{CTRF\\ classification}}}}&&Sd-Irr&11&		26&		3&		82&		619
\end{tabular}


\skpln
performance characteristics 
\skpln

\begin{tabular}{cccccc}
\hline
&E&LBS&S0-Sa&Sab-Scd&Sd-Irr\\
TPR &   70.7 &   75.6 &   63.6 &   56.4 &   88.9\\
PPV &   66.5 &   72.0 &   68.3 &   69.6 &   83.5\\
F  &    68.5 &   73.7 &   65.9 &   62.3 &   86.2
\\ \hline
\end{tabular}

\end{center}

\caption{
As for Table \ref{T:SVM}, but for the CTRF classifier.
\label{T:CTRF}}
\end{table}
%


%
\begin{table}
\begin{center}

Visual classification
\skpln

\begin{tabular}{cccccccc}
&&&  E&LBS&S0-Sa&Sab-Scd&Sd-Irr\\
&&E&117&		13&		28&		7	&	4 \\
&&LBS&9&		130&		3	&	11	&	27 \\
&&S0-Sa&27&		0&		110	&	23	&	2 \\
&&Sab-Scd&12&		3&		27	&	162	&	38\\
\rotatebox{90}{\parbox{3mm}{\multirow{3}{*}{ \ct{0.3cm}{NN\\ classification}}}}&&Sd-Irr&9	&	34&		8	&	77&		625
\end{tabular}


\skpln
performance characteristics 
\skpln

\begin{tabular}{cccccc}
\hline
&E&LBS&S0-Sa&Sab-Scd&Sd-Irr\\
TPR &   67.2 &   72.2 &   62.5 &   57.9 &   89.8\\
PPV &   69.2 &   72.2 &   67.9 &   66.9 &   83.0\\
F   &   68.2 &   72.2 &   65.1 &   62.1 &   86.3
\\ \hline
\end{tabular}

\end{center}

\caption{
As for Table \ref{T:SVM}, but for the NN classifier.
\label{T:NN}}
\end{table}
%


%
\begin{table}
\begin{center}

Visual classification

\begin{tabular}{ccccc}
&&&&\\
&&&Spheroid&Disk\\
&&&&\\
&&Spheroid&450&73\\
&&&&\\
&&Disk&80&903\\
&&&&\\
\rotatebox{90}{\parbox{3mm}{\multirow{3}{*}{ \ct{0.1cm}{binary CTRF\\ classification}}}}&&&
\end{tabular}


\skpln
performance characteristics 
\skpln

\begin{tabular}{ccc}
\hline
&Spheroid&Disk\\
TPR &   84.9 &   92.5\\
PPV &   86.0 &   91.9\\
F   &   85.5 &   92.2
\\ \hline
\end{tabular}

\end{center}

\caption{
As for Table \ref{T:SVM}, but for the binary CTRF classifier.
\label{T:bCTRF}}
\end{table}

For Tables \ref{T:SVM}--\ref{T:NN}, the left diagonal represents the objects that are correctly classified by the respective classifiers. For.eg., in Table \ref{T:SVM}, 122, 138, 112, 149 and 621 objects which were visually classified as E, LBS, S0-Sa, Sab-Scd and Sd-Irr were correctly classified by the SVM classifier. The other columns show how many of the objects were classified into which other galaxy types. The same format is followed in all the confusion matrices.

A general trend that is observed for all classifiers is that the 'misclassifications' by the classifiers are mostly from neighbouring classes. For e.g., in table \ref{T:SVM}, most of the misclassifications by the SVM classifier of the visual E galaxies are as type S0-Sa. Another interesting inference is that galaxies visually classified as classes LBS and Sd-Irr are frequently confused with each other by all four classifiers. This hints at a possible similarity in properties between these galaxy types.

The confusion matrix of the binary CTRF classifier shown in Table \ref{T:bCTRF} is similar to that of the multi-class classifiers. The actual and predicted classifications are represented by the columns and rows respectively. 450 spheroid-dominated and 903 disk-dominated objects are classified correctly by the binary classifier while the misclassifications are for 80 and 73 objects respectively. 

The PPV for the corresponding classes gives a measure of classification error by showing how exact the classifier is. For e.g. in Table \ref{T:SVM}, in the case of type Sab-Scd, while the SVM classifier only positively classifies $53.2\%$ of the time, there is a probability that when it does, it is $67.1\%$ correct. This measure depends heavily on how balanced the data set is, i.e., if there are more objects of a certain galaxy class in the data sample, that particular galaxy type will have a higher value of PPV. This can be seen clearly in the case of galaxy type Sd-Irr for all the classifiers. It can also be observed in the case of the binary CTRF classifier, for which the data set is more balanced than for multi-class classification, there is a subsequent increase in the PPV of spheroid-dominated objects (which is still the minority class).

The F-score represents the balance between the precision and recall for the classifier. For an unbalanced data set such as ours, the classifier could, in theory, get a higher accuracy rate just by choosing a majority class. In such cases, an F-score is often used to choose an optimum classifier, by choosing one that has consistently high F-scores for all the classes. In the case of the four algorithms considered in this study, that classifier is CTRF as can be seen for both the binary and multi-class classifications.

The CT algorithm is observed to be the lowest grossing method over the entire sample, with an average accuracy of 69.0\%. The other three methods, CTRF, SVM and NN have comparable values for classification accuracy at 76.2\%, 75.8\% and 76.0\% respectively. This leads us to conclude that perhaps the choice of parameters is a more important factor in classification accuracy rather than the choice of algorithms. Figure \ref{venn} represents the classification efficiencies of these three methods by GAMA Hubble type and for the entire test set. Here, CTRF, SVM and NN algorithms are represented by green, pink and blue respectively. The number of objects that are classified `correctly' by each method is shown in brackets next to the algorithm labels. The number of objects not classified `correctly' by any of the three algorithms is given in the top left corner while the total number of visual Hubble types is given in the top right corner. As can be seen in the case of each individual visual Hubble type and in the total test set (panel 6), the overall performance of the CTRF classifier is slightly better than the other two. Based on these results, we recommend the CTRF classifier for further use in astrophysical practice. Even though the improvement in classification accuracy is marginal, CTRF has a simpler mathematical structure. The CTRF machine learnt classifications will be our primary automatic classifications used for further analysis below.

Figures \ref{image1}-\ref{image5} show several example postage stamp images of different galaxy types from our test set. The postage stamps span an area of $3\times$Kron radius of each galaxy and are ordered according to their stellar masses (low-mass galaxies at the top, high-mass galaxies at the bottom). Classifications for different statistical learning algorithms are overlaid on the top right corner of these images in the order SVM, CT, CTRF and NN. As can be seen, the majority of machine learnt classifications agree well with their visual Hubble type, however, there are instances where one or more algorithms classify a galaxy as something different from its visual classification. All four algorithms are in agreement with each other in 1040 out of the 1506 galaxies in our test set. And out of these 1040 objects, 143 (i.e. $\sim10\%$ of the total test set) differ from the respective visual classification. This `unanimous disagreement' occurs with varying frequency for the different morphological types\footnote{All the numbers quoted here (and henceforth in the same context) are percentages on the total test set.}: $\sim9\%$ for type E, $\sim9\%$ for type LBS, $\sim14\%$ for type S0-Sa, $\sim21\%$ for type Sab-Scd and $\sim4\%$ for type Sd-Irr. This phenomenon could be due to two reasons, (1) the visual classification might be inaccurate and, based on the parameters that were used for training, the galaxy belongs to a different class, or, (2) some vital information to classify this galaxy is missing, i.e., the given parameters are not sufficient. Figure \ref{undis} shows a few examples of galaxies that exhibit this phenomenon. Further analysis of this interesting occurrence is required to explore why a host of machine learning algorithms may consistently agree with one another yet disagree with the human eye.

\subsection{Analysis : CTRF classifier}

Figures \ref{xbox} and \ref{tprz} represent the TPRs obtained by the CTRF classifier as a function of the total stellar mass and redshift respectively for the galaxies in our test set. In both cases the errors are calculated using the {\tt aqbeta} function from the astro library in R \citep{Cameron2011}. This estimates the confidence intervals from quantiles of a beta distribution fit to the data, and is especially suited for small to intermediate data samples. 

In Figure~\ref{xbox}, the TPRs obtained by the CTRF classifier are plotted against the total stellar masses of the galaxies from our test set. The first panel represents all galaxies while the distributions of distinct GAMA HTs are plotted in the subsequent panels (see legend). We find that the accuracy in classification decreases as the total stellar mass increases. This becomes evident in the extreme mass trends observed for HTs S0-Sa and Sab-Scd. In the case of elliptical galaxies (type 1, E), the TPR values seem to be increasing after a dip at $\log_{10} M_\odot \sim 10.5$. This seems to be a real rather than a statistical effect, as the bin centred at $log_{10}M_\odot =10.5$ has more objects in it than the one centred at $log_{10}M_\odot =11$. For type Sd-Irr, the success rate drops significantly from $ \sim 90\%$ at low mass to $ \sim 30\%$ at $\log_{10}  M_\odot > 10$. It seems that the algorithm finds it increasingly difficult to classify type Sd-Irr at higher masses, however, we note that the very low number statistics for this population in this mass regime (both in training and test sets), as evidenced by the relatively large error bars could also be a contributing factor. This trend holds true for type LBS as well. \citet{Moffett2016} notes that types LBS and Sd-Irr together account for only about 10\% of the total stellar mass density of the parent sample, and that their frequencies drop to nearly zero above the mass range $log_{10}M_\odot =10.0$. The reason for the decrease in TPR values in the case of early and intermediate-type spirals isn't clear at this time, but may be related to the increasingly apparent complexity of structure in galaxies of these types at higher mass regimes. 


Figure \ref{tprz} is a similar representation of the TPRs with the redshifts of all the galaxies in the test set along the x-axis. The first panel represents all the galaxies in our test set while the succeeding panels represent the different HTs (see legend). For the total sample, the trend is to be expected, considering that we have attempted to choose redshift independent parameters. However, we observe varying trends along the sub-populations. The trend for each HT sub-population is similarly consistent with a flat relation with redshift, with the notable exception of type Sab-Scd, for which the TPR is lower at low redshifts and goes on to increase at higher redshifts. This may be due to the fact that local galaxies are better resolved than distant galaxies, and therefore the automated algorithms may be having a harder time processing the extra structural data. The apparent angular scale from $z=0.02$ to $z=0.06$ decreases by a factor of $\sim3$, which has the effect of blurring stellar populations within the galaxies.

Figures \ref{histn1}-\ref{histn15} show the location of galaxies in the S\'ersic index --- $g-i$ colour plane with each figure representing a different visual Hubble type morphology. Data point types and colours represent the morphological types assigned to each galaxy by the CTRF classifier. The marginal histograms represent the distributions of $g-i$ colour (top) and S\'ersic index (right) for the visual and CTRF classifications. The efficiency of classification by the CTRF classifier for different Hubble types can be visually inspected from these histograms.  

Figure \ref{histn1} shows all visually classified elliptical galaxies in the S\'ersic index vs $g-i$ colour plane. Most of the objects for which the classifier is unable reproduce the visual classification are determined to be early-type spirals (S0-Sa). The objects that have been classified by the CTRF classifier as S0-Sa are all redward of the main population, whilst other types are scattered in the blue low S\'ersic index tail of the E distribution. One reason for this could be the potential systematic misclassification of face-on red S0 galaxies as ellipticals. If true, our machine learning algorithm may provide a robust automated means by which we could apply corrections to currently existing visual morphological datasets to address the issue of E/S0 confusion. Another reason for this `spheroid-disk tension' between the human eye and the automated algorithms could be the presence of disky elliptical `ES' (\citealt{Liller1966}; \citealt{Graham2016}; \citealt{Savorgnan2016}) class with intermediate-disks in our sample. It could also be a wider `red disk detection' issue, however, we note that the S\'ersic indices for many of these objects are of the order of $n\sim4$ which indicates spheroid-dominated systems.

Figure \ref{histn2} shows objects that are visually classified as little blue spheroids (type 2, LBS, represented as green squares). The instances where the CTRF classifier is not in agreement with the visual classifications are represented by the other colours and points in the scatter plot. In general, most of the objects which were not found to be LBS by the CTRF method have been classified as late-type spirals \& irregulars, except towards the redder end of the scatter plot, where they have been classified as elliptical galaxies. We note that in the visual classification of this particular type, the `blue colour' was a secondary characteristic, the objects were primarily classified on the basis of their shape and size. 

Figure \ref{histn1112} shows objects visually classified as early-type spiral galaxies (type 1112, S0-Sa, barred and unbarred, represented as black diamonds). The CTRF classifier's classifications that do not agree with the visual morphology are almost equally divided between ellipticals (red circles) and intermediate-type spirals (purple triangles). They seem to be uniformly distributed in Sersic index space, while there appears to be some dependence in $g-i$ colour, with the objects classified as ellipticals clustered in an area redder than the objects that are classified as intermediate-type spirals. Classification as intermediate-type spiral follows a trend observed by \citet{owens1996}, in that differentiating between neighbouring classes of galaxies such as these is more difficult than differentiating between non-neighbouring classes. The population of elliptical galaxies we find might be an indicator that the human eye is fallible when classifying this type of galaxy. Very few objects are classified as late-type spirals \& irregulars or little blue spheroids (mostly at the bluer end). 

Figure \ref{histn1314} shows objects that are visually classified as intermediate-type spirals (type 1314, Sab-Scd, purple triangles). In most instances where the CTRF classifier disagrees with the visual classification, it classifies objects as late-type spirals \& irregulars. However, at the redder and higher S\'ersic index end, some objects are classified as early-type spirals. This is also the galaxy type for which the classifiers of the machine learning algorithms that we have applied disagree the most with visual classifications.

Figure \ref{histn15} shows objects that are visually classified as late-type spirals \& irregulars (type 15, Sd-Irr, represented as blue triangles pointing down). For this particular galaxy type, all four machine learning algorithms have a high agreement rate with the visual classifications ($ >80\%$). As is shown, the disagreements are evenly divided between types LBS and intermediate-type spirals, while there are a few objects classified as ellipticals. The classifications as LBS and ellipticals could be an indication that these objects may have more in common with early-type galaxies than is currently conceived. The classifications as intermediate-type spirals are likely due to the \citet{owens1996} observations mentioned previously.

\subsection{Impact of chosen parameters on the CTRF classifier} \label{pcapar}

\begin{table*}
\caption{Results of parameter sensitivity test on the CTRF algorithm in percentages are shown in panel 1. In panel 2 similar results for redundant parameters according to the PCA performed in Section \ref{pcasec} (Figure \ref{prin}) are shown. The results for the CTRF classifier from the original run are shown in panel 3.}
\label{fsall}
\resizebox{\textwidth}{!}
{
\begin{tabular}{lllllll}
\hline
\\
\ct{2cm}{Parameter \\removed}&\ct {2cm}{E \\1} & \ct{2cm}{LBS \\2} &\ct{2cm}{S0-Sa \\ 1112} & \ct{2cm}{Sab-Scd\\1314} & \ct{2cm}{Sd-Irr \\ 15} & All\\
\\
\hline
\\
S\'ersic index &67.8 &   76.7 &   58.5 &   55.7 &   87.9&    74.8\\
\ct{3cm}{Kron radius\\ in kpc\\(semi-minor axis)}&69.0&    71.7&    62.5&    55.7&    88.7&    75.2\\
\ct{3cm}{Half-light radius\\ in kpc }&65.5 &   68.9&    60.8&    56.8 &   90.5 &   75.3\\
\ct{3cm}{Kron radius\\ in kpc\\(semi-major axis)}&67.8&    70.6 &   61.4&    56.8 &   89.8&    75.5\\
Ellipticity&66.7&    74.4 &   63.1&    57.1 &   88.8 &   75.6\\
Mass-to-light ratio&69.5 &   71.7 &   62.5 &   57.5 &   89.1  &  75.8\\
$g-i$ colour & 67.2  &  76.7&    63.6&    56.4 &   88.8 &   76.0\\
Stellar mass&70.7 &   71.7&    64.2&    57.5&    88.8&    76.0\\
$u-r$ colour&70.7&    74.4 &   62.5 &   56.8 &   89.0&    76.0\\
Absolute magnitude&71.8 &   73.3  &  61.4&    58.2&    90.0&    76.5\\
\\
\hline
\\
\ct{3cm}{Mass-to-light ratio\\ \& $g-i$ colour}& 68.4 &  75.0 &   63.7 &   60.0 &   89.1&    76.6\\
\ct{3cm}{Mass-to-light ratio \\ \& $u-r$ colour}&67.8 &   75.6 &   60.8 &   57.9 &   89.7 &   76.2\\
\ct{3cm}{$u-r$ colour \\ \& $g-i$ colour }&69.5 &   76.1 &   61.4  &  60.4 &   89.4 &   76.8\\
\\
\hline
\\
All chosen parameters&$70.7^{+3.2}_{-3.7}$ & $75.6^{+2.9}_{-3.5}$ & $63.6^{+3.5}_{-3.8}$ & $56.4^{+2.9}_{-3.0}$ & $88.9^{+1.1}_{-1.3}$ & $76.2^{+1.1}_{-1.1}$\\
\\
\hline

\end{tabular}}

\end{table*}

We perform a sensitivity test to ascertain the impact of each parameter on the classification process of our CTRF algorithm. In order to achieve this, we remove all the parameters mentioned in the upper panel of Table \ref{features} one by one, and obtain the TPRs, re-training the CTRF classifier in each instance. The results of this are shown in Table \ref{fsall}.

The removal of S\'ersic index lowers the overall rate of accuracy the most, by almost $1.4\%$. All other increases and decreases from the overall TPR caused by the removal of parameters are within the error limits defined in Table \ref{T_TPRs}. The only parameter whose removal causes an increase in the overall TPR is absolute magnitude, by $0.3\%$.  This indicates that for the total data sample, S\'ersic index is the parameter that contributes most to the classification process by the CTRF algorithm. This, however, does not hold true for the individual Hubble types.

Removal of $u-r$ colour and stellar mass does not affect the classification in the case of elliptical galaxies. Absolute magnitude and mass-to-light ratio have an almost similar effect on the TPR values, albeit in different directions. When absolute magnitude is removed, the TPR value increases by $1.1\%$ and when mass-to-light ratio is removed, the value decreases by $1.2\%$. The parameters for which the accuracy falls outside the error bars are ellipticity and half-light radius. 

In the case of LBS galaxies, the parameters that affect the classification process the most are half-light radius, Kron radius (semi-major and semi-minor), mass-to-light ratio and stellar mass. The parameters that have a similar effect on the classification rate are Kron radius (semi-minor axis), mass-to-light ratio and stellar mass, a decrease by $\sim 4\%$. The decrease in TPR values is drastic in the case of both half-light radius and Kron radius (semi-major axis), $\sim7\%$ and $\sim5\%$ respectively.

For early-type spiral galaxies, the changes in TPR are within the error bars except in the case of S\'ersic index. When S\'ersic index is removed prior to training the classifier, the accuracy drops by $\sim5\%$. The effects caused by the absence of Kron radius (semi-minor), mass-to-light ratio and $u-r$ colour are analogous, a decrease of $\sim 1\%$. Same is the case with Kron radius (semi-major) and absolute magnitude, by $\sim2\%$. When $g-i$ colour is excluded from the process, the TPR values remain the same as that from the original run.

The change in accuracy for intermediate-type spirals after removing the parameters one by one, are all within the error limits of the values from Table \ref{T_TPRs}. As in the case of early-type spirals, removing 
$g-i$ colour has no effect on the original TPR values. S\'ersic index and Kron radius (semi-minor) contribute to a decrease in TPR values by $\sim0.7\%$ each; Kron radius (semi-major), half-light radius and $u-r$ colour to an increase by $\sim0.4\%$ each; and mass-to-light ratio and stellar mass to an increase by $\sim1\%$ each. Removing absolute magnitude seems to matter the most, by increasing the accuracy by $\sim2\%$.

The changes in TPR in the case of late-type spirals \& irregulars are mostly within the error bars of the original results, except in the case of half-light radius where it increases by $\sim2\%$, which seems to have the most impact on classification accuracy as well. Ellipticity, $g-i$ colour and stellar mass have a similar effect on the TPR values (decrease by $0.1\%$). $u-r$ colour seems to have a similar impact on the classifier's performance for this galaxy class, an increase of the TPR by $0.1\%$.

In the PCA we performed (represented in Figure \ref{prin}), ellipticity was found to be the parameter which contained the least variability. But as can be seen from Table \ref{fsall}, while it might not be the most important parameter overall, it has a significant impact in the classification accuracies of individual Hubble types, especially elliptical galaxies. The TPR of ellipticals fall by $4\%$ when this parameter is removed. 

Also represented in Figure \ref{prin} is the redundancy of the parameters, mass-to-light ratio, $g-i$ and $u-r$ colours. We also explore here, the impact on the classification accuracies when these parameters are removed two at a time.These results are represented in the second panel of Table \ref{fsall}.

 When mass-to-light ratio and $g-i$ colour are removed, there is a marginal increase in the overall TPR value, to $76.6\%$. This increase is reflected in the individual Hubble types, S0-Sa, Sab-Scd and Sd-Irr. The accuracies take a consequent dip in the case of types E and LBS. 
 
 The removal of mass-to-light ratio and $u-r$ colour does not make a significant overall impact, with the TPR value remaining the same as that of the original run, at $76.2\%$. Among the individual Hubble types, the accuracy of LBS remains unchanged while that of types E and S0-Sa decrease. The individual TPRs of types Sab-Scd and Sd-Irr reflect marginal increases.
 
Removing $g-i$ and $u-r$ colours resulted in an increase in the overall TPR value, to $76.8\%$. This increase was contributed by the increases in the TPRs of galaxy types LBS, Sab-Scd and Sd-Irr. The accuracies of types E and S0-Sa was found to drop marginally. 

The slight increases and decreases in the TPR values when the parameters are removed one by one are largely within the error margins defined for the TPRs from the original run and therefore are not deemed significant. Similar is the case when redundancies in parameters are removed.\footnote{It is interesting to see that the TPR values for Sab-Scd, the class that performs the worst during classification by all our algorithms, experience significant increases when the redundant parameters are removed. However, since this doesn't make a noteworthy change in the overall rate of accuracy, we have decided to overlook this improvement and keep the parameter set as is.} Therefore we conclude that, while the individual Hubble types might be sensitive to certain parameters more than the others, all parameters contribute to some extent in the overall classification process of the CTRF algorithm.

\subsection{CTRF classifier for binary classification}

\begin{table*}
\caption{Panel 1 shows the results of parameter sensitivity test performed with the binary CTRF classifier. The results with all chosen parameters (Table \ref{features}) are shown in panel 2.}
\label{fsbinary}
\begin{tabular}{llll}
\hline

\\
\ct{0.1cm}{Parameter\\removed} & \ct{1.5cm}{Spheroid\\-dominated}&\ct{1.5cm}{ Disk\\-dominated}& All \\
\\
\hline
\\
\ct{3cm}{Half-light\\ radius in kpc}& 78.7 & 92.1 & 87.4 \\
\vspace{0.1cm}
S\'ersic index&82.5 & 90.4& 87.6 \\
\vspace{0.1cm}
\ct{2cm}{Kron radius \\in kpc\\(semi-major axis)}& 82.6 &91.4 &88.3\\
\vspace{0.1cm}
\ct{2cm}{Kron radius \\ in kpc\\(semi-minor axis)}&83.8 &91.2 &88.6\\
\vspace{0.1cm}
Mass-to-light ratio&83.6 &91.4 &88.7 \\
\vspace{0.1cm}
Stellar mass&83.0 &92.0 &88.8\\
\vspace{0.1cm}
Ellipticity& 84.9 &91.3 &89.0\\
\vspace{0.1cm}
$g-i$ colour& 84.3 &91.6 &89.0\\
\vspace{0.1cm}
Absolute magnitude&84.8 &91.6 &89.2\\
$u-r$ colour& 83.8 &92.1 &89.2\\
\\
\hline
\\
\ct{2cm}{All chosen \\parameters}&$84.9^{+1.4}_{-1.7}$ &$92.5^{+0.8}_{-0.9}$ &$89.8^{+0.7}_{-0.8}$\\
\\
\hline
\end{tabular}
\end{table*}

With the same training, test and parameter sets that we have employed in multi-class classification, we constructed a binary CTRF classifier with two classes, spheroid-dominated and disk-dominated\footnote{We use this terminology based on the visual classification of the data set. Since lenticular galaxies are gathered under the same umbrella as Sa-type galaxies, an early-type to late-type galaxy split would involve re-classifying the entire visual sample, which is beyond the scope of this work.}. The galaxies which were visually classified as ellipticals (type 1, E), little blue spheroids (type 2, LBS) and early-type spirals (type 1112, S0-Sa) were considered as spheroid-dominated while the intermediate-type spirals (type 1314, Sab-Scd) and late-type spirals \& irregulars (type 15, Sd-Irr) were considered as disk-dominated. 

This binary CTRF classifier returned a total success ratio of $89.8\%^{+0.7}_{-0.8}$ with individual TPRs of $84.9\%^{+1.4}_{-1.7}$ and $92.5\%^{+0.8}_{-0.9}$ for the spheroid-dominated and disk-dominated classes respectively. This significant increase from the original CTRF classifier's TPRs proves that as the number of classes into which classification is made increases, the classification accuracy decreases. This might also be directly related to the size of the data set, and how well each class is represented in the training set. 

Similar to the analysis in Section \ref{pcapar}, we also explored the impact the different parameters might have on the classification performance of the classifier constructed by the CTRF algorithm. The results of this are given in Table \ref{fsbinary}.

Removing half-light radius from the parameter set used for training and testing the CTRF algorithm seems to be the have the most impact on the performance of the binary CTRF classifier. While the overall success rate drops by $2.46\%$, the values for spheroid-dominated and disk-dominated systems fall by $\sim6\%$ and $\sim0.4\%$ respectively. This points at the greater significance of half-light radius in the classification of spheroid-dominated galaxies rather than the disk-dominated ones. This is in agreement with the results represented in Table \ref{fsall} in which the classification accuracies fall consistently for these three classes (E, LBS and S0-Sa) in the case of multi-class classification.

Ellipticity \& $g-i$ colour and absolute magnitude \& $u-r$ colour seem to have similar overall effect on the classification process, drops by $\sim0.8\%$ and $\sim0.6\%$ for the respective pairs. The fluctuations in the TPR values are most significant in the case of ellipticity for disk-dominated systems. The entire contribution to the change in TPR while ellipticity is removed as a classifying criterion comes from disk dominated systems. This is a very interesting development because, in the case of multi-class CTRF classification discussed in Section \ref{pcapar}, ellipticity is one of the parameters that cause the TPR to decrease for all three galaxy types collectively called as spheroid-dominated. This might indicate cross-contamination between these three galaxy types in the visually classified sample which confuses the classifier.

The accuracy rates (both overall and individual) fall beyond the error margins when parameters such as S\'ersic index, Kron radii (major and minor axes), mass-to-light ratios and stellar mass are removed. According to this study, the parameters that influence our CTRF algorithm the most are half-light radius, S\'ersic index, Kron radii, mass-to-light ratio and stellar mass.

\begin{figure}
\includegraphics[width=8cm]{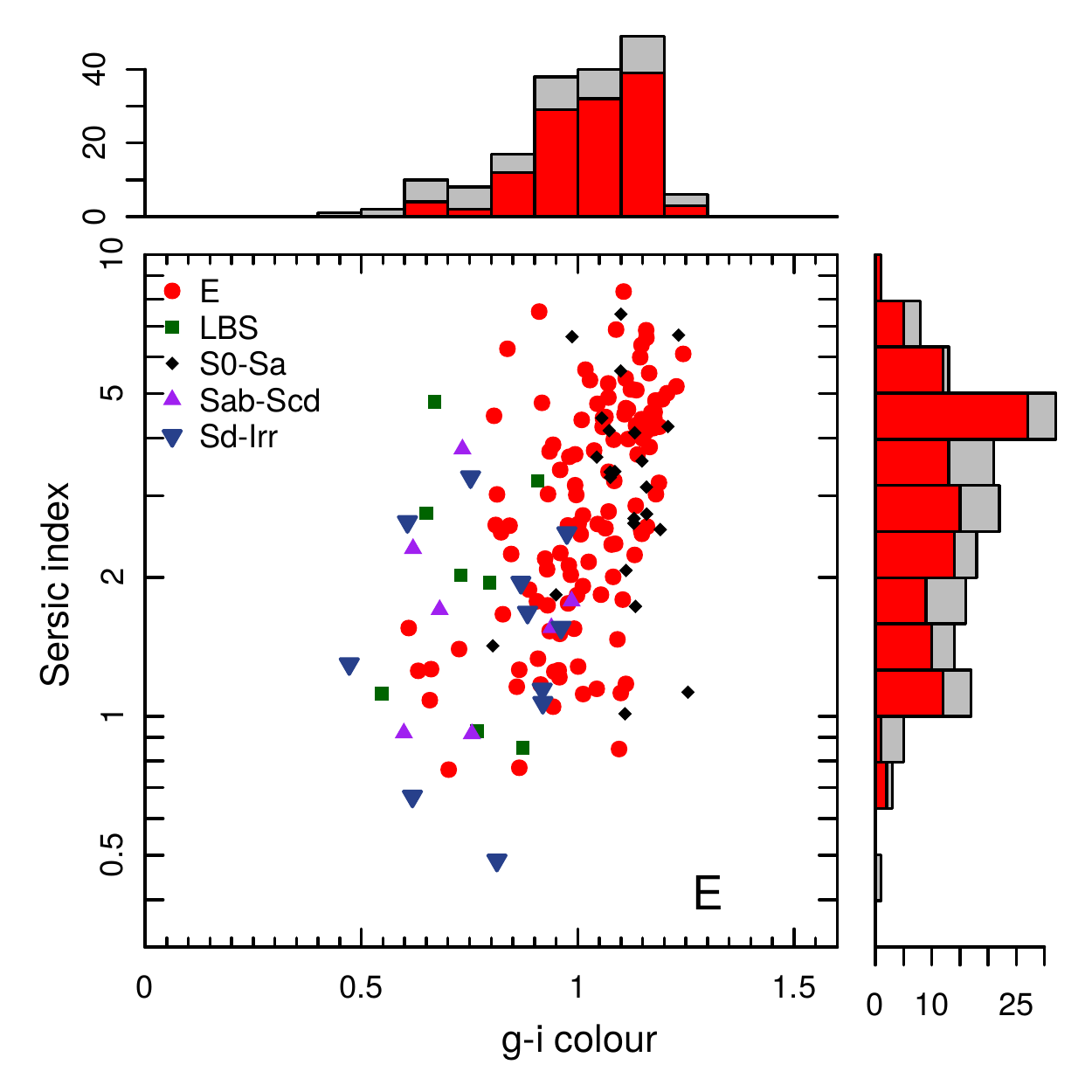}
\caption{Scatter plot with marginal histograms showing all visually classified elliptical (type 1 , E) galaxies in S\'ersic index and $g-i$ colour space. Data point colours and types vary according to their CTRF classification, as indicated by the inset legend. Marginal histograms show the distribution for all (grey) and visually classified elliptical (red) galaxies. }
\label{histn1}
\end{figure}

\begin{figure}
\includegraphics[width=8cm]{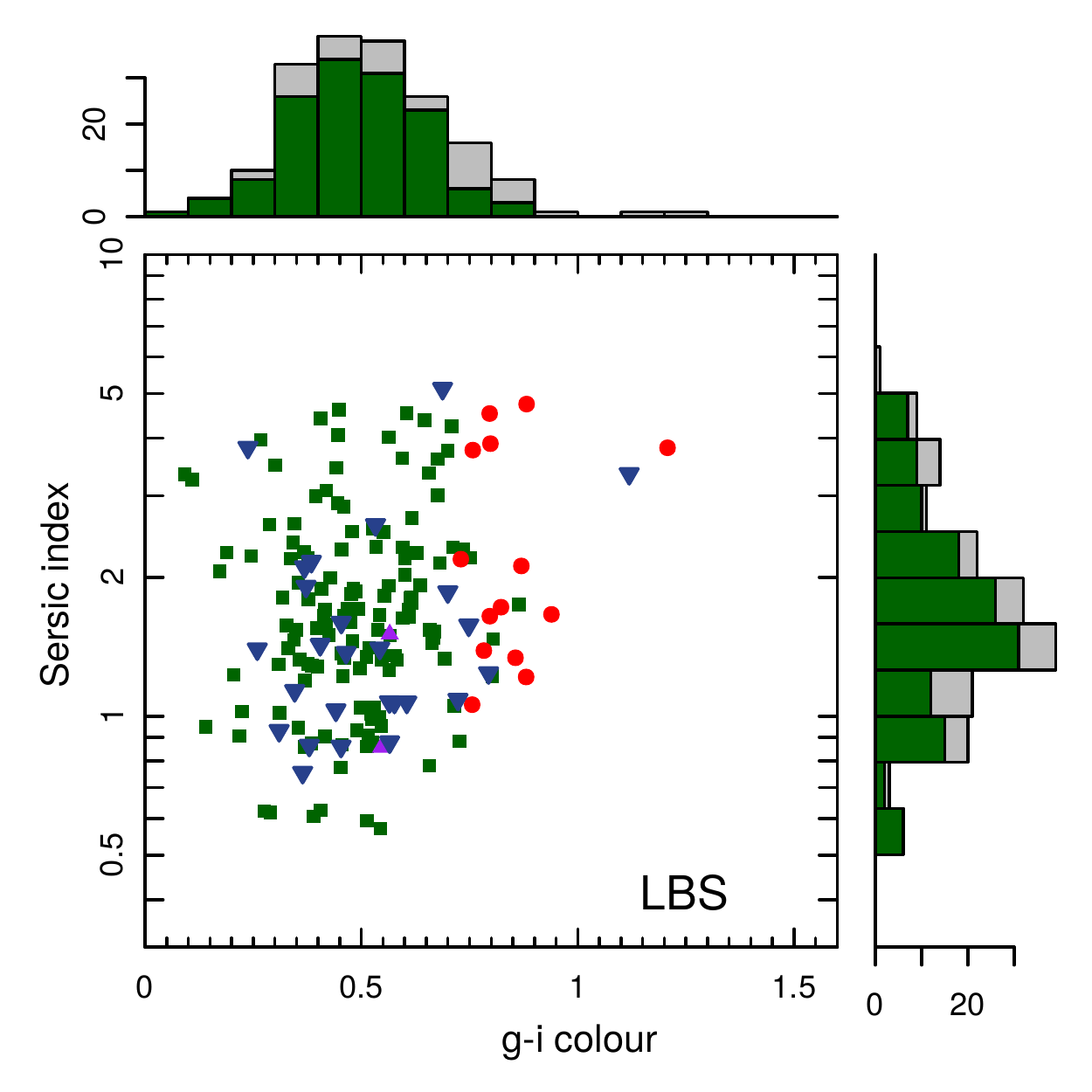}
\caption{As Figure \ref{histn1}, for Little Blue Spheroids (type 2, LBS). }
\label{histn2}
\end{figure}

\begin{figure}
\includegraphics[width=8cm]{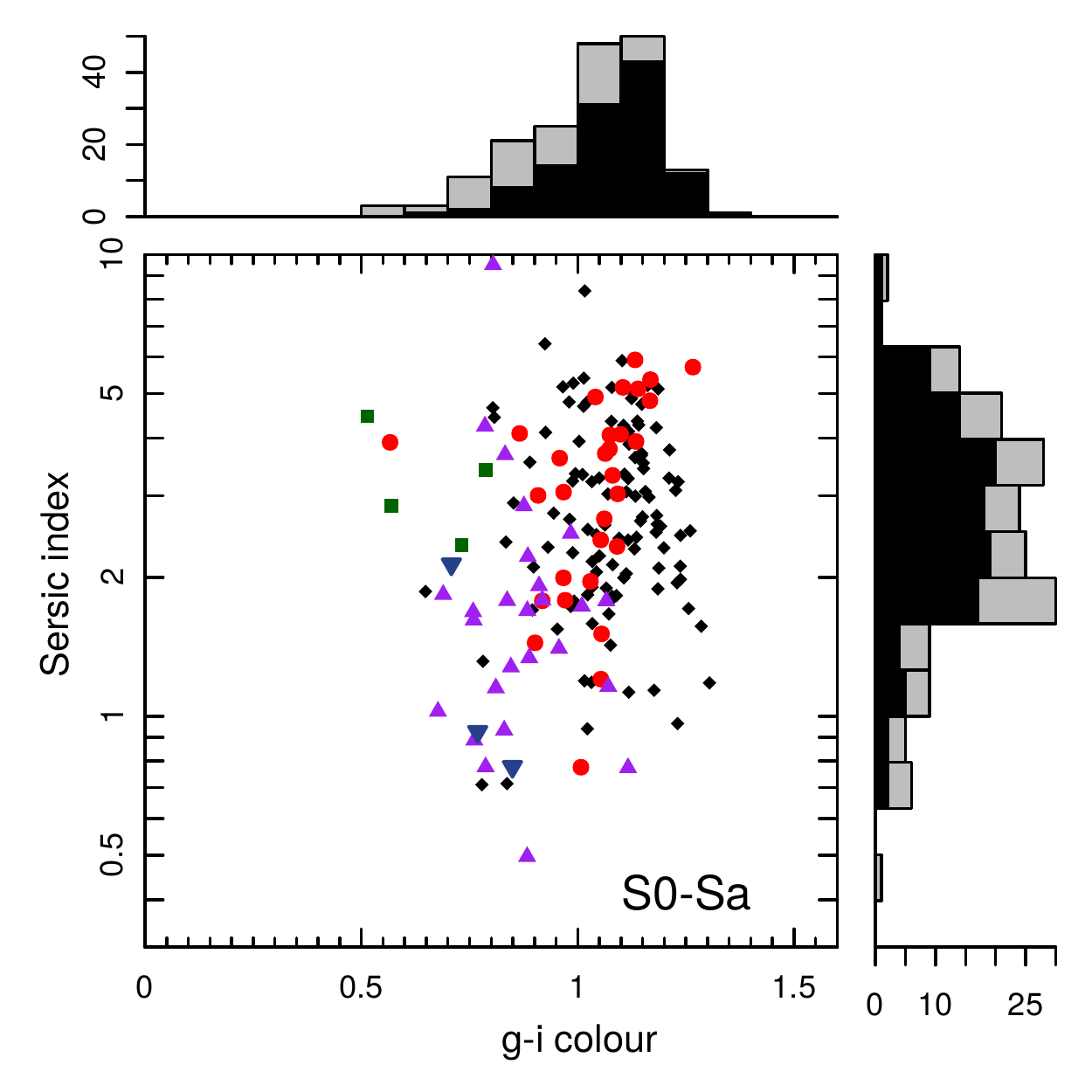}
\caption{As Figure \ref{histn1}, for early type spirals (type 1112, S0-Sa, barred and unbarred). }
\label{histn1112}
\end{figure}

\begin{figure}
\includegraphics[width=8cm]{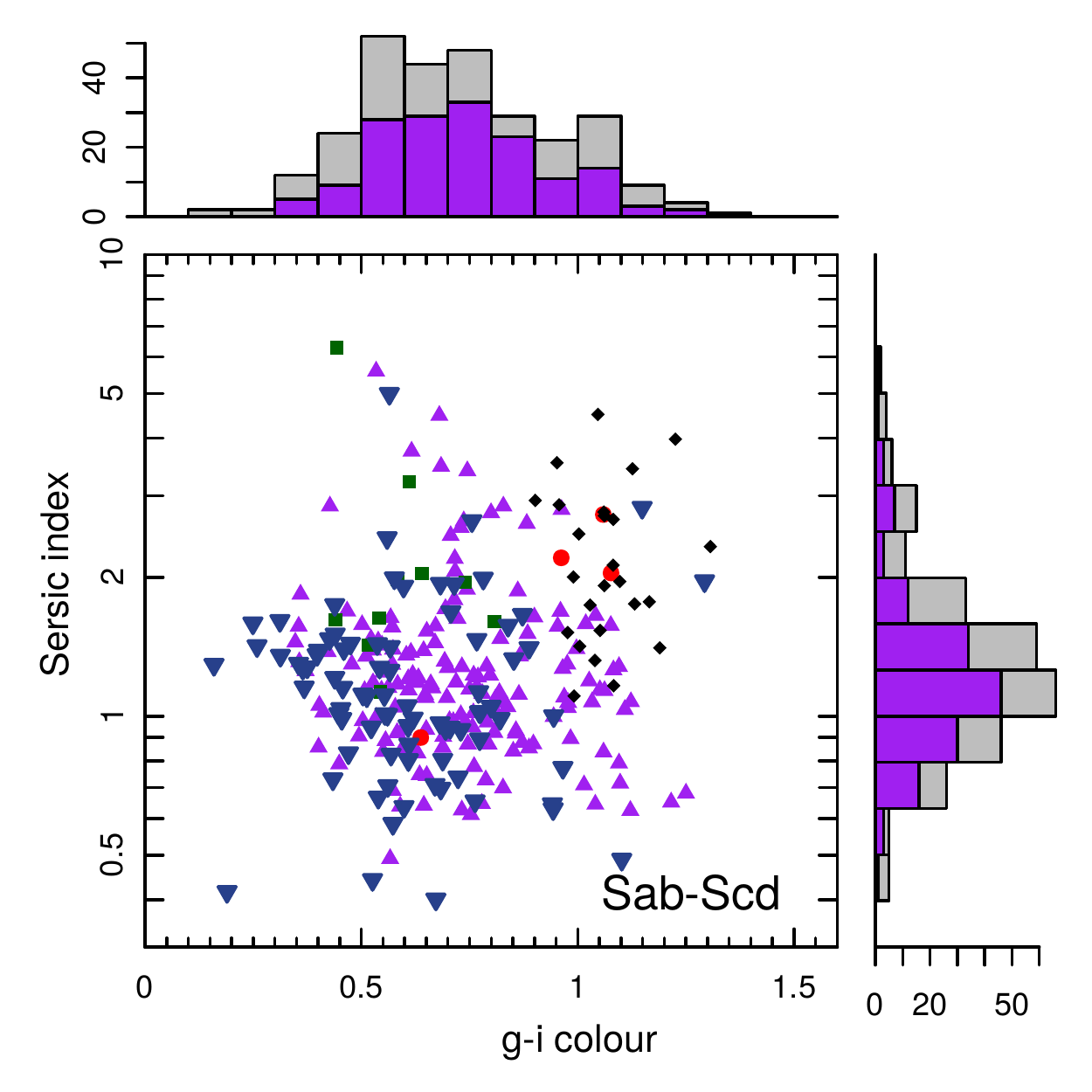}
\caption{As Figure \ref{histn1}, for intermediate type spirals (type 1314, Sab-Scd, barred and unbarred).}
\label{histn1314}
\end{figure}

\begin{figure}
\includegraphics[width=8cm]{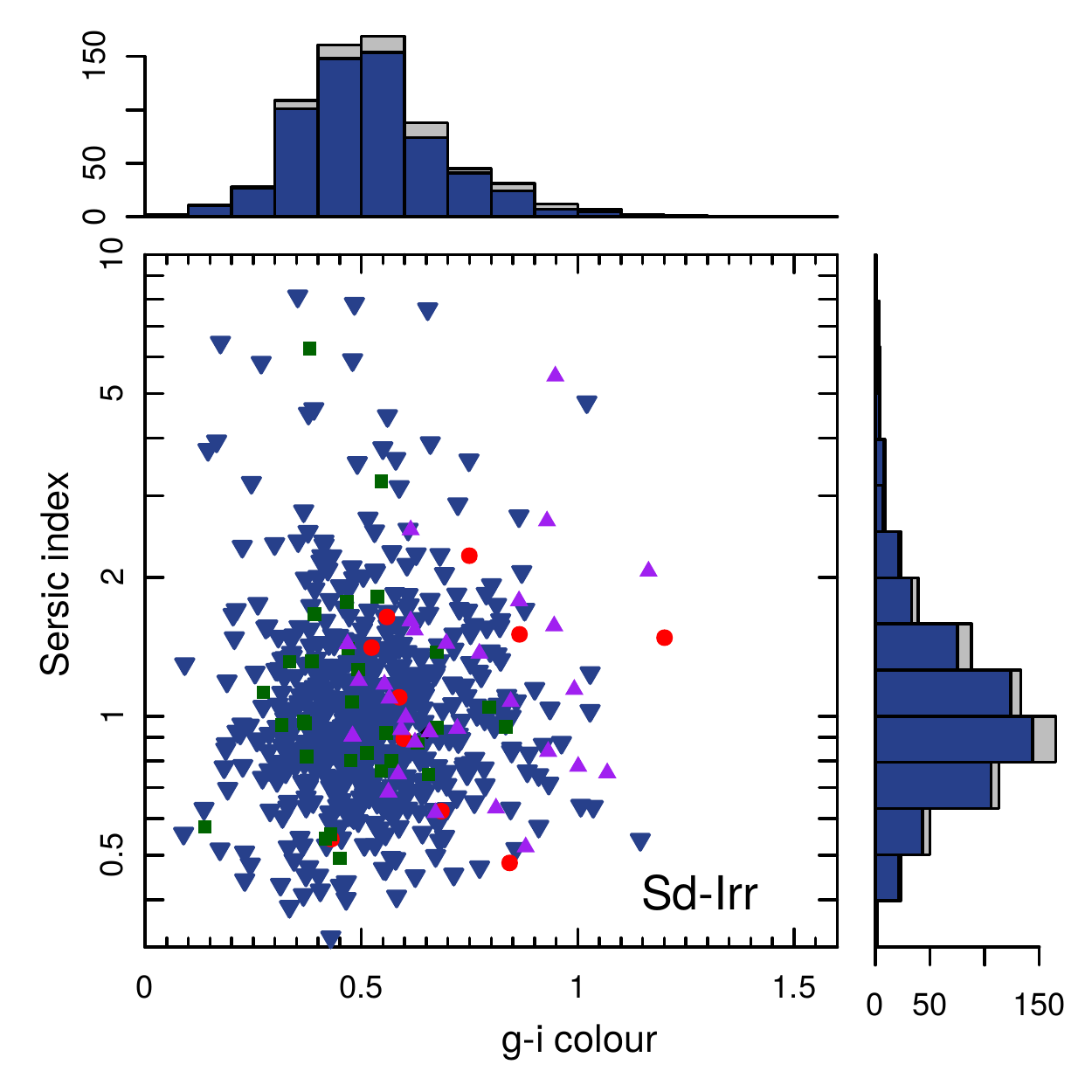}
\caption{As Figure \ref{histn1}, for late type spirals \& irregulars (type 15, Sd-Irr). }
\label{histn15}
\end{figure}

\section{Discussion} \label{disc}

\begin{table*}
\caption{Summary of the results from this study (top) alongside results from several other studies from the literature using a variety of statistical learning methods (bottom).}
\label{summ}
\resizebox{\textwidth}{!}
{
\begin{tabular}{llllllll}
\hline
\ct{2cm}{Statistical \\ Learning \\Method} & Total sample & Training set& Test set & \ct{2cm}{Number of \\ classes} &Dimensions&Accuracy & Reference \\

\hline
\hspace{2cm}\ldelim\}{10}{20pt} & & & & &  \hspace{1.6cm}\rdelim\{{10}{20pt}&  \\
SVM &  &   &  &  & & 75.8\% & \\
\\

& 7528  &6022 (80\%) &1506 (20\%) &5  &10 & &  \\

NN &  &  & &  &  & 76.0\% & Results from our work \\
\\
 
CT &  & & & &  & 69.0\% & \\
\\
CTRF &  & & & &  & 76.2\% & \\

\\
\hline
\\
\\SVM & $\sim1500$ & 500 (33\%) & 1000 (67\%) & 2 (early-type, late-type) & 12&80\% & \citet{Huertas-Company2007}\\
\\NN & $\sim$1,000,000&$\sim$75,000 (7.5\%)&$\sim$925,000 (92.5\%)&\ct{2cm}{3 (early-type, spirals, point sources/artefacts)}&12&90\%&\citet{banerji2010}\\
\\Oblique CT & 5217 & $\sim$4174 (80\%) &$\sim$1043 (20\%)&5 (E, S0, Sa+Sb, Sc+Sd, Irr)&13&63\%&\citet{owens1996}\\
 \\ \ct{2cm}{Three CT \\algorithms\\including CTRF}&75,000&67,500 (90\%)&7500 (10\%)&3 (ellipticals, spirals, unknown)&13&96.2\%&\citet{gauci2010}\\
\\ConvNet & 58,000&47,700 ($\sim82\%$)&5000 ($\sim9\%$) &  5 (probablities$^\dagger$)& run on images & $\sim$99\% & \citet{Huertas-Company2015}\\
    &    &   & \ct{2cm}{5300 ($\sim$9\%) used for \\real-time evaluation\\during training}  &   &   &   & \citet{Dieleman2015}\\ 
    
\hline
\end{tabular} }

\raggedright$^\dagger$Probabilities for each galaxy having a disk or a spheroid, being a point source, having an irregularity or being unclassifiable are the outputs.
\end{table*}

In this section, we discuss in greater detail our previously recovered results. To begin, we note that type 15 (Sd-Irr galaxies), account for almost 50\% of our test set, and the associated TPR success values are above 80\% for all considered automated classification methods. This could indicate one of three scenarios; (1) As the percentage of objects in a certain class increases, the accuracy of classification increases as well, (2) the algorithms that we tested are more effective in classifying a particular Hubble type (type 15 in our case) using the parameters that we have prescribed or (3) the human classifications may be biased towards being able to more accurately classify Sd-Irr type galaxies. 

The first scenario is not generally supported by our own results. The TPR values for type 1314 are consistently low across all 4 considered methods and yet it is the second most populous type in both our training and test sets. This warrants additional analysis in future works; by testing the codes on larger data samples and by fine-tuning the classification algorithms by introducing techniques such as cross-validation.

 As to the second scenario, the successful utilisation of our adopted functions are directly linked to our choice of parameters. It may be that one or more of the parameters that we have chosen are more effective in classifying certain Hubble types while falling short in others. For example, the complexity in the structure of the galaxy might not be well defined by the parameters that we have chosen. As can be seen in Table \ref{T_TPRs}, the TPR values are considerably higher for single component systems such as ellipticals (type 1, E) and late-type spirals/Irregulars (type 15, Sd-Irr) compared to multi-component systems such as early and intermediate type spirals (types 1112, S0-Sa and 1314, Sab-Scd respectively).

All four algorithms are in agreement with each other in 1040 out of the 1506 galaxies in our test set. And out of these 1040 objects, 143 (i.e. $\sim10\%$ of the total test set) disagree with the classification by visual inspection. Of these,  $\sim9\%$ are ellipticals, $\sim9\%$ are LBS, $\sim14\%$ are early-type spirals, $\sim21\%$ are intermediate-type spirals and $\sim4\%$ are late-type spirals \& irregulars. These are illustrated in Figure \ref{undis}. There seems to be an element of symmetry in this occurrence. For instance, as can be seen from the figure, no objects that have been visually classified as S0-Sa are machine classified unanimously as Sd-Irr, this pattern holds true in converse as well. But this isn't always the case. No visual LBS galaxies have been unanimously machine classified as S0-Sa objects but one visual S0-Sa galaxy has been machine classified unanimously as LBS. This, along with the possibility of unanimous disagreement being a potential indicator of human error in classification by visual inspection are interesting paths to follow in future works that extend this study. 

 When we train a machine, for e.g., to classify galaxies (our case) based on visual classifications, what we essentially do is train it to reproduce our classification strategy, replete with our human biases. For instance, if, beyond a certain redshift, the human eye is ineffective in distinguishing between certain classes of galaxies, the data set that we apply to the algorithms will reflect the same bias. Therefore, we propose that the disagreement between the machine and the visual classifications could be due to one of two reasons, (1) the visual classification is inaccurate, and based on the values of the parameters used to train and test the algorithms, the galaxy belongs to one of the other classes, or (2) the parameters do not sufficiently characterise what we see while classifying by eye. 
 
In Figures \ref{histn1}, \ref{histn2} and \ref{histn15} it can be seen that the CTRF method replicates the visual classification to a greater extent than in Figures \ref{histn1112} and \ref{histn1314}. This leads us to speculate that our algorithms in their present configuration might be more effective in classifying single component systems such as ellipticals and late-type spirals rather than multi-component systems like early and intermediate-type spirals.

One of the methods that we have used in our work is SVM with a tree structure. With this approach, the accuracy obtained on our entire test set is 75.8\%. The accuracies for the different HTs are represented in Table \ref{T_TPRs}. This value seems encouraging when we compare our results to \citet{Huertas-Company2007}, who also used an SVM approach in their work to obtain morphological classification to a sample of 1500 galaxies from the SDSS (500 to train, 1000 to test). Their method was a generalisation of C-A system using non linear SVM boundaries with 12 dimensions. The mean accuracy of the method was $\sim80$\%. We note that the \citet{Huertas-Company2007} method only classifies galaxies into early and late types while our algorithm classifies galaxies into five distinct morphological types, which may explain why their success ratio is $\sim4\%$ higher than ours.

In our NN method, we reproduce the classifications learned on the training set to an accuracy of 76.0\% on the test set. \citet{banerji2010} applied artificial neural networks to a sample of almost one million objects from the SDSS previously classified by human eye by volunteers as part of the Galaxy Zoo project. Their training set consisted of 75000 objects, classifying the test set into three morphological classes (early-types, spirals and point sources/artefacts) with 12 parameters. The accuracy of their approach was close to 90\%. Considering that our training set and test sets are much smaller compared to \citet{banerji2010} and that we use a larger range of classification types, our value of 76.0\% is highly promising.

Our CT algorithm uses classification (decision) trees to attain morphological classification with an accuracy of 69.0\% on our entire test set. The size of the data set and the number of classification types for the method of \citet{owens1996} is comparable to our own. They use a sample of 5217 galaxies from the ESO-LV catalogue \citep{lv1989} using 13 parameters to discern between five morphological types (ellipticals, lenticulars, early-type spirals, late-type spirals and Irregulars). With a five-fold cross validation on their approach they achieved an average accuracy of 63\% on a test set which amounted to 1/5th of the whole set. They have compared their results with \citet{slss1992} which applied an artificial neural network approach to the same data with an accuracy of 64.1\% and \citet{lv1989} whose automated classifier reproduced classifications to an accuracy of 56.3\%. We note however, that \citet{slss1992} have used $\sim30\%$ of their total data sample as the training set and ~70\% as the test set in contrast to our method of adopting a larger training set and smaller test set as detailed in Section \ref{s_dprep}. The improvement of 69.0\% accuracy that we observe is undoubtedly due to this reason. Furthermore, we have ~2000 more objects in the data sample which will influence the classification accuracy.

Among our methods the CTRF algorithm which employs a random forest of 100 trees was found to have an accuracy of 76.2\%. This method has a marginal but encouraging higher accuracy among all four methods that we have tested. \citet{gauci2010} performed a comparison of different classification tree algorithms to a data set of 75000 objects from the SDSS previously classified by the Galaxy Zoo project. The algorithms of CART, C4.5 and Random Forest (RF) are tested with a ten-fold cross validation technique where, in each run, nine subsets of the data are used for training and one for testing. The success rate was 97.33\% for an  RF algorithm with 50 trees and 96.2\% over all the methods. However, \citet{gauci2010} have only 3 classification types (elliptical, spiral and unknown morphology) compared with 5 in this study. 

We trained a binary CTRF classifier that classifies our data sample to spheroid-dominated and disk-dominated systems. For this, we consider galaxy types E, LBS and S0-Sa as spheroid-dominated and galaxy types Sab-Scd and Sd-Irr as disk-dominated. The overall accuracy rate for this classifier is $\sim90\%$ with individual TPRs for spheroid-dominated and disk-dominated systems to be $\sim85\%$ and $\sim93\%$ respectively.

The results from our binary CTRF classifier has clarified certain aspects about the effectiveness of our overall study. The results indicate that the number of data types into which the classification is done is a very important criterion for accuracy. There is an increase of almost $14\%$ overall accuracy when the number of types changed from  5 to 2. It is conceivable that the size of the data set and how comprehensively the different galaxy types are represented in the training and test sets play a role in the performance accuracy as well. This can be seen in the higher accuracy of the disk-dominated galaxies which make up $\sim67\%$ of the total data set. So a way to address the decrease in accuracy as the galaxy types increase might be to increase the size of the data set accordingly.

To facilitate future studies and to aid in comparison with other works (see Table \ref{summ}), the machine learning algorithms employed in this study have not been significantly modified beyond their default setups as detailed in Section~\ref{s_meths}. There are several avenues that could be pursued in order to make them more precise. Applying the SVM method for multi-class classification using error-correcting output codes is one such approach (\citealt{DieBak95}). There are indications in literature that this technique could be more accurate than the tree structure that we have considered in this work. Assigning probabilities to our classifications rather than binary values may be a useful tool to see the effectiveness of the classification process. \citet{owens1996} posits that differentiating between neighbouring classes of galaxies (for e.g., types 1112 and 1314 in our sample) is more complicated than differentiating between non-neighbouring classes of galaxies. By analysing the probabilities assigned to each class by the classification algorithms and manual examination, it might be possible to define criteria or introduce parameters that provide a more robust delineation between neighbouring galaxy types. Introducing Principal Component Analysis (PCA) as a means to choose a robust set of parameters and extensive error analysis of the parameters that we have chosen are other interesting prospects, allowing for the introduction (e.g. some measure of asymmetry) or elimination of parameters (e.g. ellipticity) which do not seem to be vital in predicting morphology. The methods that we have chosen construct classifiers with different mathematical structures.Therefore each constructed classifier may capture different aspects of the ideal classifier effectively. Using a combination of classifiers constructed using different statistical learning methods  may give rise to a new classifier with better accuracy (and closer to an ideal classifier) than each classifier taken individually (\citealt{ChePX15,KriPPT16}). The design of appropriate combination strategies is another avenue that we may explore in the future.

\section{Conclusion}
In this study, we have used the statistical machine learning algorithms of Support Vector Machines (SVM), Classification Trees (CT), Classification Trees with Random Forests (CTRF) and Neural Networks (NN) to carry out morphological classifications for 7528 galaxies from the Galaxy and Mass Assembly (GAMA) survey. These galaxies were previously visually classified independently by three classifier teams and the majority vote has been included in the GAMA catalogue. The algorithms are trained on a set of 6022 objects (80\% of the data set) using 10 distance independent parameters. These algorithms are subsequently tested on the remaining 20\% of the data set (1506 objects) to classify them into five galaxy types: elliptical (type 1, E), little blue spheroid (type 2, LBS), early-type spirals (type 1112, S0-SBa), intermediate-type spirals (type 1314, Sab-SBcd) and late-type spirals \& irregulars (type 15, Sd-Irr). We draw the following conclusions from our study.

\begin{enumerate}[i]
\item The success rates on the entire test set are 69.0\%, 76.2\%, 75.8\% and 76.0\% for the CT, CTRF, SVM and NN algorithms respectively. While the performance of the SVM, CTRF and NN algorithms are very similar, the CTRF algorithm has a marginally better success rate and a simpler mathematical structure. We therefore recommend this algorithm to provide robust, automated Hubble type classifications when applied to future extragalactic surveys.

\item Our algorithms have a greater success rate in the case of single component systems such as ellipticals, late-type spirals \& irregular galaxies. This is especially clear when we look at the success rate of type 15 galaxies (Sd-Irr). They form 47\% of our entire sample. The success rates of all four algorithms are above 80\% for this galaxy type and close to 90\% for CTRF, SVM and NN algorithms. 

\item We find that the success rates decrease with increasing stellar mass. This trend seems drastic in the case of HTs S0-Sa, Sab-Scd and Sd-Irr. This apparent phenomenon warrants further investigation. 

\item We do not find a universal trend in the success rates with respect to redshift, however, we find that there is some redshift dependence within each galaxy type. This is especially apparent in the case of type Sab-Scd, for which, the success rates are lower at lower redshifts and increase towards higher redshifts. 

\item In the cases where all 4 machine learning algorithms agree with each other, they disagree with the visual classification  $\sim10\%$ of the time, with $\sim9\%$ being ellipticals, $\sim9\%$ LBS, $\sim14\%$  S0-Sa, $\sim21\%$ Sab-Scd and $\sim4\%$ Sd-Irr. These unanimous disagreement fractions could be a potential indicator for human error in visual classifications. Further exploration of this is an interesting path to investigate for future work.

\item When we decrease the number of galaxy types into which classification is done, the accuracy of classification increases considerably. Our binary CTRF classifier achieved an overall accuracy of $89.8\%$ with the spheroid-dominated and disk-dominated classes achieving accuracies of $84.9\%$ and $92.5\%$ respectively. This hints that a way to cope with the decrease in classification accuracy as the galaxy types increase might be to use larger data sets.

\item There are many possible avenues to pursue following from this study. These include introducing analysis methods such as PCA or cross-validation to create a robust dataset of input features, foregoing the SVM tree structure in favour of error-correcting codes, and using an ensemble of classifiers constructed using different statistical learning methods.

\end{enumerate}

\section*{Acknowledgements}

GAMA is a joint European-Australasian project based around a spectroscopic
campaign using the Anglo-Australian Telescope. The GAMA input catalogue
is based on data taken from the Sloan Digital Sky Survey and the UKIRT
Infrared Deep Sky Survey. Complementary imaging of the GAMA regions
is being obtained by a number of independent survey programmes including
GALEX MIS, VST KiDS, VISTA VIKING, WISE, Herschel-ATLAS, GMRT and
ASKAP providing UV to radio coverage. GAMA is funded by the STFC (UK),
the ARC (Australia), the AAO, and the participating institutions.
The GAMA website is http://www.gama-survey.org/.

Sergiy Pereverzyev Jr. gratefully acknowledges the support of the Austrian Science Fund (FWF): project P 29514-N32.


\bibliography{ref.bib}
\bibliographystyle{mn2e}

\appendix

\section{Appendix : Methods in detail}

\subsection{Support Vector Machines} \label{svmapp}

\begin{figure}
\begin{center}
\includegraphics[width=8cm]{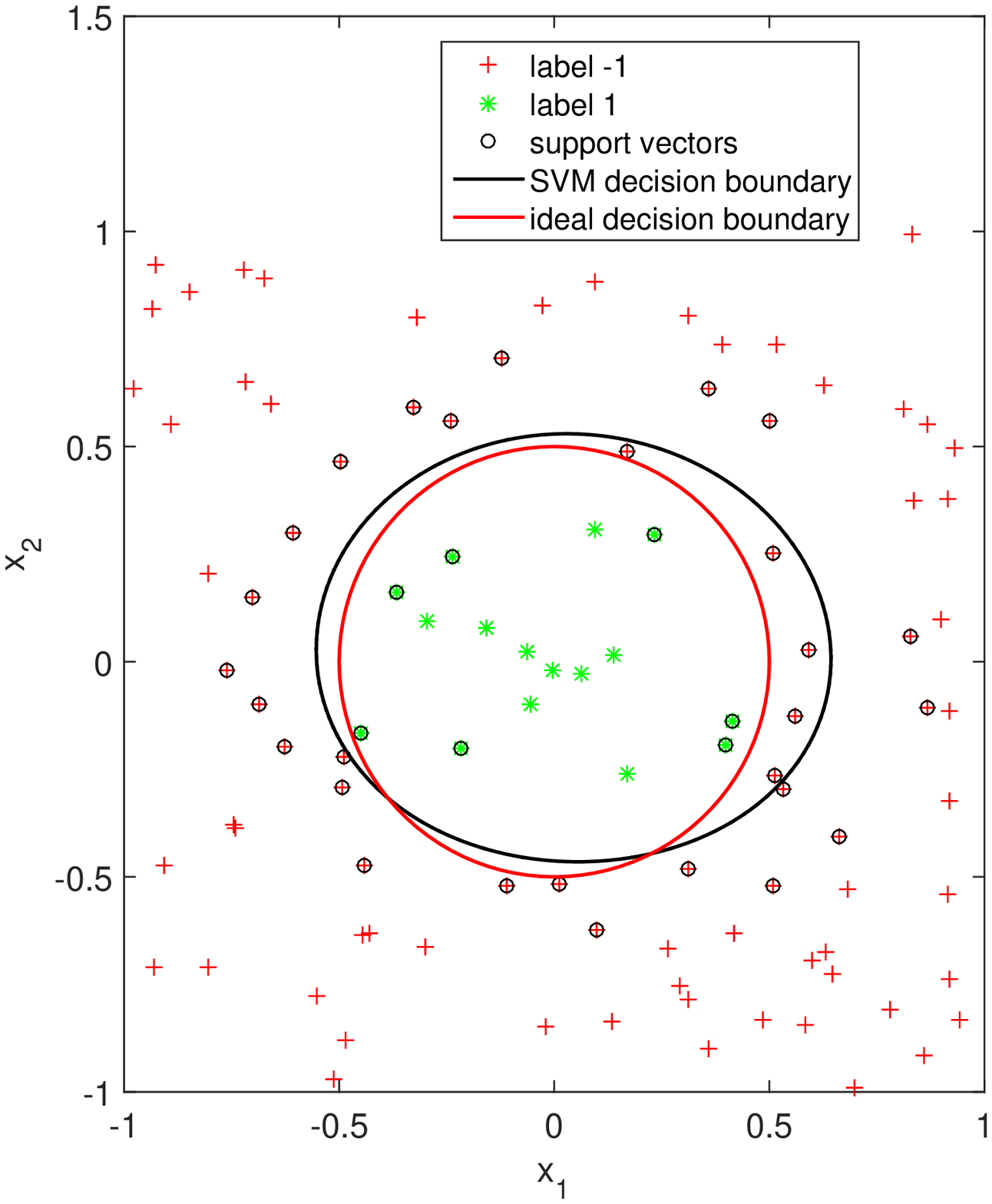}
\caption{Illustrative example for the SVM method in the case of two features $x_1,x_2$. The feature vector belongs to the class $1$ if it is inside the red curve, which is the ideal decision boundary. Otherwise, the feature vector is in the class $-1$. The black
curve is the decision boundary constructed by the SVM method using the training data in the picture. The corresponding constructed SVM classifier assigns the feature vectors inside this black curve to class $1$, and outside the black curve to class $-1$. The training feature vectors that are additionally marked by a small surrounding circle are the support vectors.}
\label{FM_SVMd}
\end{center}
\end{figure}

In SVM, the constructed classifier $f$ is of the following form:
\begin{equation}
	f\kl{ \xv } = \sign\kl{g\kl{ \xv }}
	= \sign\kl{b + \sum\limits_{i=1}^{N} \alpha_i K\kl{  \xv, \xv_i }}
\label{SVMrp}
\end{equation}
where $K$ is a positive definite kernel\footnote{Recall that a symmetric function $K\colon \R^p\times\R^p\to \R$ (here symmetric means that
$ K\kl{ \bar\xv_1,\bar\xv_2 } = K\kl{ \bar\xv_2,\bar\xv_1 } $ for any $\bar\xv_1,\bar\xv_2\in\R^p$)
is called {\it positive definite kernel} if for any $m\in\N$ and any distinct
$\bar\xv_1,\dots,\bar\xv_m \in \R^p$, the $m\times m$ matrix $\bar\Km$ with entries
$\bar\Km_{ij} = K\kl{ \bar\xv_i,\bar\xv_j } $ is positive definite.
}, and
$b,\alpha_i,\; i=1,2,\dots,N$ are certain coefficients from $\R$. We set
$\alpv:= \kl{ \alpha_1,\alpha_2, \dots, \alpha_N }^\top $.

The coefficients $b$, $\alpv$ are chosen as the solution of the following minimisation problem:
\begin{equation}
\label{SVMmp}
				\sum\limits_{i=1}^N \kl{ 1-z_i g\kl{ \xv_i } }_+
				+ \frac{\lambda}{2}  \alpv^\top \Km \alpv
				\longrightarrow
				\min\limits_{b,\alpv},
\end{equation}
where $(a)_+ := \max (0,a)$, $\Km$ is the $N\times N$ kernel matrix with entries $\Km_{ij} = K\kl{ \xv_i,\xv_j }$,
and $\lambda >0$ is a penalty parameter. Note that the minimisation problem in Equation~\eqref{SVMmp} is convex, and therefore, various methods of convex optimisation
(e.g., \citealt{BoyVan04}) can be used to solve it. We employ the 
{\it sequential minimal optimisation} method, which is suggested in MATLAB as the standard method to solve this. 
The first term in Equation~\eqref{SVMmp} measures the closeness of
$f\kl{\xv_i} = \sign\kl{ g\kl{ \xv_i } }$ to $z_i$, i.e. it tells us how well the classification is
predicted on the training set, while the second term penalises coefficients in $\alpv$, 
and $\lambda$ gives a tradeoff between the two terms.
We take the default value for $\lambda$, which is
$\lambda=1$. 

Due to the nature of the 
function $(\cdot)_+$ in Equation~\eqref{SVMmp}, many values of $\alpha_i$ are equal to $0$.
Therefore, in the representation in Equation~\eqref{SVMrp}, the linear combination involves functions from a subset of
$\Set{  K\kl{  \cdot,\xv_i  },\; i=1,2,\dots,N  }$, and the corresponding $\xv_i$ are called {\it support vectors}.

The kernel that we have chosen is the {\it Gaussian Radial Basis Function} given as: 
\begin{equation}
\label{GRBF}
   K\kl{ \xv,\xv' } = \exp\kl{   -\frac{  \norm{  \xv - \xv'  }^2   }{   2\sigma^2   }   },
\end{equation}
where $\sigma$ is the scaling factor, whose default value we have retained as $\sigma=1$.

The SVM method is illustrated
by an example in Figure~\ref{FM_SVMd}.
In this example, the feature vector
is $\kl{ x_1,x_2 }  \in [-1,1]\times [-1,1]$ and $D:= \Set{ \kl{ x_1,x_2 } |  x_1^2 +x_2^2 \leq \kl{ 1/2 }^2   }$ is a  disk with its centre as $(0,0)$ and radius $1/2$. The ideal classifier $f^*$ assigns the feature vector
to class $1$ if it belongs to $D$, and to class $-1$ otherwise.
We generate a training set $\zv$ that consists of $100$ feature vectors
$ \xv_i = \kl{  x_{1,i},x_{2,i}  } $. Features $x_{1,i}$, $x_{2,i}$ are randomly sampled using the uniform distribution over $[-1,1]$. 
The classes 
for the feature vectors $\xv_i$ in the training set
are determined using
the  ideal classifier $f^*$.
Then, we construct the function $g$ using the described SVM method with kernel Equation~\eqref{GRBF} and $\sigma=1$.

The red curve is the boundary of $D$, which can be called as the {\it ideal decision boundary}. The black curve consists of feature vectors $\xv$ for which $g(\xv)=0$. This curve is called the {\it SVM decision boundary}. For the feature vectors inside this curve, we have $g(\xv)>0$, and therefore, these feature vectors will be classified by the constructed SVM classifier $f$ as class $1$.
The other feature vectors satisfy the condition $g(\xv)<0$, and therefore, they are assigned by $f$ to class $-1$. The training feature vectors inside the small circles are the support vectors. 

In general, the SVM decision boundary may have an arbitrary shape, and it may also consist of several closed curves.
The support vectors are located near the SVM decision boundary, in a way supporting and defining its shape.

\subsection{Classification Trees with hyper-rectangular partitions} \label{ctapp}

In the CT method, the feature space is split into the rectangular partitions by a recursive binary method.  First, it is split into two regions,
$\Set{  \xv\in \R^p |  x_i<s  }$ and $\Set{  \xv\in \R^p |  x_i \geq s  }$ using a selected feature $x_i$ and a split
point $s$. Then, one or both of these regions are split similarly into two more regions. this process continues until a certain
stopping condition is fulfilled.

An example of the above described partition for two features $\kl{  x_1, x_2  }$ with values in the unit square
is presented in Figure~\ref{CTpart2}. In this example, the first split is made at $x_1 = s_1$. Then, the region
$\Set{  \xv\in \R^p |  x_1<s_1  }$ is split at $x_2=s_2$, and the region
$\Set{  \xv\in \R^p |  x_1\geq s_1  }$ at $x_1=s_3$. In the end, the region
$\Set{  \xv\in \R^p |  x_1\geq s_3  }$ is split at $x_2=s_4$. Thus the partition of the feature space into five rectangular regions
$R_1$, $R_2$, $\dots,$ $R_5$ shown in Figure~\ref{CTpart2} is obtained.

The nodes of the CT are split based on the 
{\it impurity measure} of the node. We represent the region that corresponds to the node $t$ as  $R_t$.  Let
$N_t:= \# \Set{  \xv_i\in R_t  }$ denote the number of training feature vectors in $R_t$. The mathematical
notation $\# R$ is used for the number of elements of a set $R$. We further define
$p_k(t):= \# \Set{  \xv_i\in R_t | y_i = k }  /  N_t  $
as the proportion of the training feature vectors in the node $t$ (or, which is the same, in the region $R_t$) that belong to class $k$. Impurity measure $I(t)$ of the node $t$ is a function of the proportions $p_k(t)$. It tells us how even the distribution
of the feature vectors in the node $t$ are over the classes. It has a maximum value
when the feature vectors are distributed evenly over the classes in the node $t$, i.e. when $p_k(t) = 1/T$,
$k=1,2,\dots, K$. In contrast, when the node $t$ contains feature vectors only from one class, say class $\ell$, i.e.
when $p_{\ell}(t)=1$, and $p_k(t)=0$, $k\neq \ell$, then the impurity measure $I(t)$ has a minimal value, and the node
is called {\it pure}. As the impurity measure, we consider the Gini index :
$I(t) = 1 - \sum_{k=1}^T p_k^2(t)$.

The goal of the node splitting is to obtain new nodes with smaller impurity measures. This is achieved by defining a characteristic called
impurity gain, and the splitting is then done such that this gain is maximized. Let $P\kl{ R_t } = N_t /N $ denote the proportion of the training feature vectors in the node $t$. Consider a particular splitting candidate
of the node $t$, i.e. a particular splitting feature and a split point, and denote the corresponding new left node as $t_1$ and the
new right node as $t_2$. Then, the impurity gain is defined as :
\begin{equation}\label{imgain}
   \Delta I = P\kl{ R_t } I\kl{t}  - P\kl{ R_{t_1} } I\kl{t_1} - P\kl{ R_{t_2} } I\kl{t_2},
\end{equation}
and then, the splitting candidate for which this impurity gain is maximum is chosen.

There is a finite number of splitting candidates. For each feature $x_q$, $q=1,\dots,p$, possible
splitting points are obtained from the training data by sorting $x_{i,q}$ in the ascending order.
Note that $x_{i,q}$ is the $q$-th component of the feature vector $\xv_i$, and those feature vectors $\xv_i$
are considered that belong to the splitting node. Then, the maximisation of the impurity gain (Equation~\ref{imgain}) is done by checking through all possible splitting candidates.

\subsection{Classification Trees with Random Forest} \label{ctrfapp}

Using random sampling {\it with replacement}, $B$ random samples $\zv_b$, $b=1,\dots,B$ of the training set $\zv$ are created.  Consider the set
$\bar\zv = \Set{ 1,2,\dots,10 }$ with 10 elements. Then, 4 random samples of $\bar\zv$ that are made
using the random sampling with replacement of 10 elements from $\bar\zv$ can be, for example, the following:

$$
\begin{aligned}
\bar\zv_1 &= (3,4,9,6,3,8,4,9,3,10),
\\
\bar\zv_2 &= (2,8,4,7,5,5,2,4,7,8),
\\
\bar\zv_3 &= (1,1,2,4,10,5,6,2,8,10),
\\
\bar\zv_4 &= (2,3,8,4,7,6,6,7,2,8).
\end{aligned}
$$

On each training sample $\zv_b$, a CT classifier $f_b$ is trained using a modified CT learning algorithm. In this modified algorithm,
at each node split, possible splitting features are taken from a random sample of all used features. Typically, these random samples
contain $\sqrt{p}$ (rounded down) features (p is the number of features, Section \ref{classprob}).
The resulting CTRF classifier assigns a classification to a feature vector $\xv$ using the majority vote
of the constructed CT classifiers $\Set{ f_b,\; b=1,\dots,B }$. 

\subsection{Single hidden layer feed forward Neural Networks} \label{nnapp}

The arrows
in the network diagram, Figure \ref{FM_NN} indicate the dependence between network units, and this dependence is modeled as :
$$
\begin{aligned}
  w_m & = g_1 \kl{  \alpha_{0m} + \alpv_m^\top \xv  },\; m=1,2,\dots,M, \\
  \bar v_k & = \beta_{0k} + \betv_k^\top \wv  ,\; k=1,2,\dots,T,\\
  v_k & = g_{2,k} \kl{ \bar\vv } =: f_k\kl{ \xv }  ,\; k=1,2,\dots,T,
\end{aligned}
$$
where $\wv = \kl{ w_1,w_2,\dots, w_M }^{\top}  $, and $\bar\vv = \kl{ \bar v_1, \bar v_2,\dots,\bar v_T }^{\top}  $.
The numbers $\alpha_{0m}$, $\beta_{0k}$ and vectors $\alpv_m\in\R^p$,   $\betv_k\in\R^M$ are model parameters called weights. The complete set of these weights is denoted by $\thtv$.
The functions $g_1$ and $g_{2,k}$ are called transfer functions. For $g_1$, we take the tan-sigmoid
transfer function:

$$  g_1(s) = \frac{\kl{  \exp(s) - \exp(-s)  } } {  \kl{  \exp(s) + \exp(-s)}   }, $$

and for $g_{2,k}$, we take the softmax transfer function,
$$
  g_{2,k}\kl{  \bar \vv  } = \frac{  \exp\kl{ \bar v_k }  }{  \sum_{\ell=1}^{T} \exp\kl{  \bar v_{\ell}  }  }.
$$
The softmax transfer function ensures that the unit values $v_k$ belong to the interval $(0,1)$ and satisfy
$\sum_{k=1}^T v_k = 1$, which allows $v_k$ to be interpreted as the probability to belong to class $k$.
The mentioned conditions on $v_k$ require the second transfer function $g_{2,k}$, in contrast
to the first transfer function $g_1$, to vary with $k$.

Once the weights $\thtv$ of the neural network are chosen, the NN classifier is defined as :
$$f\kl{ \xv } = \argmax_k f_k\kl{ \xv }$$ i.e., for any feature vector, the class with the highest probability is taken.

During the training process, the weights $\thtv$ of the neural network are tuned such that the error function $E(\thtv)$ is minimised.
The error function describes how well the NN model fits the training data. As the error function, we consider
the cross-entropy function:
$$
   E(\thtv) = -\sum\limits_{i=1}^N  \sum\limits_{k=1}^T  v_{ik} \log f_k\kl{ \xv_i },
$$
where $v_{ik}=1$ if $y_i=k$, and $v_{ik}=0$ otherwise. The minimisation of the error function can be done by gradient based methods. We use the scaled
conjugate gradient backpropagation algorithm~(\citealt{Mol93}) which is suggested in MATLAB for tuning
neural networks used for classification problems.

\end{document}

%% file: comms.tex
\newcommand{\skpln}{\vspace*{\baselineskip}}

\newlength{\ctw}\newlength{\ctwa}
\newcommand{\ct}[2]{\parbox{#1}{\vspace*{0.1cm} #2 \vspace*{0.1cm}}}

\newcommand{\kl}[1]{\left(#1\right)}

\newcommand{\norm}[1]{\left\| #1 \right\|}

\newcommand{\R}{{\mathbb R} }
\newcommand{\N}{{\mathbb N} }

\newcommand{\tm}[1]{\mathrm{#1}}

\DeclareMathOperator{\sign}{sign}
\DeclareMathOperator*{\argmax}{argmax}

\newcommand{\Km}{ \mathbf{K} }

\newcommand{\xv}{ \mathbf{x} }

\newcommand{\zv}{ \mathcal{Z} }
\newcommand{\wv}{ \mathbf{w} }
\newcommand{\vv}{ \mathbf{v} }

\newcommand{\alpv}{ \boldsymbol{\alpha} }
\newcommand{\thtv}{ \boldsymbol{\theta} }
\newcommand{\betv}{ \boldsymbol{\beta} }